\documentclass{iopjournal}
\usepackage[T1]{fontenc}
\usepackage{babel}
\usepackage{graphicx,amsmath,amssymb,color}
\usepackage[normalem]{ulem}
\usepackage{amsfonts}
\usepackage[toc,page]{appendix}
\usepackage{hyperref}
\usepackage{latexsym}
\usepackage{amsfonts}
\usepackage{algpseudocode}
\usepackage{amsthm}
\usepackage{mathrsfs}
\usepackage{color,verbatim}
\usepackage{psfrag}
\usepackage{algorithm}
\usepackage{algpseudocode}
\usepackage{rotating} 
\usepackage{bbold} 
\usepackage{multirow}
\usepackage{graphicx}

\usepackage{subcaption} 

\usepackage[svgnames]{xcolor}
\usepackage{tikz}

\usepackage{algorithm}

\algrenewcommand\algorithmicrequire{\textbf{Input:}}
\algrenewcommand\algorithmicensure{\textbf{Output:}}

\newcommand{\bra}[1]{\langle#1 |}
\newcommand{\ket}[1]{|#1 \rangle}

\newcommand{\ketbra}[2]{\vert #1 \rangle \! \langle #2 \vert}

\fancypagestyle{firstpage}{%
  \fancyhf{}%
  \fancyhead[R]{Rathore {\it et al}\ }%
  \fancyfoot[L]{\footnotesize $^*$Corresponding author: \href{mailto:omer.rathore@durham.ac.uk}{omer.rathore@durham.ac.uk}}%
  \fancyfoot[C]{\thepage}%
  \renewcommand{\headrulewidth}{0.5pt}%
  \renewcommand{\footrulewidth}{0.5pt}%
}

\date{\today}

\begin{document}

\articletype{Topical Review}
\title{Encoding strategies for quantum enhanced fluid simulations: opportunities and challenges}
\author{Omer Rathore$^{1,*}$, Alastair Basden $^1$, Nicholas Chancellor $^2$ and Halim Kusumaatmaja $^{3}$}\\
\thispagestyle{firstpage}
\affil{$^1$ Department of Physics, Durham University, Durham, DH1 3LB, UK} \\
\affil{$^2$ School of Computing, Newcastle University, Newcastle upon Tyne, NE4 5TG, UK} \\
\affil{$^3$ Institute for Multiscale Thermofluids, School of Engineering, The University of Edinburgh, Edinburgh, EH9 3FB, UK}

\begin{abstract}
Quantum computing has emerged as a powerful potential accelerator for computational fluid dynamics (CFD), but whether this promise can be realized in practice depends on how fluid information is encoded on quantum hardware. This review provides an architecture-agnostic assessment of encoding strategies for quantum-enhanced fluid simulation, focusing on the trade-offs they impose on state preparation, measurement, boundary treatment, nonlinear dynamics, and temporal evolution. We examine the principal encoding paradigms used in the literature and relate them to representative quantum algorithms for fluid simulation. Through these examples, we show that encoding choices fundamentally shape both the algorithm itself and also the practical feasibility of quantum CFD. 
For example, highly compact encodings can offer attractive asymptotic advantages but might introduce severe bottlenecks in readout, state preparation, and nonlinear processing, whereas less compact representations may simplify interactions and improve compatibility with analog and near-term hardware. 
No single encoding is universally optimal, rather the most suitable choice depends strongly on the structure of the fluid problem, the computational objective and the constraints of the target quantum platform. 
We therefore argue that encoding should be treated as a primary design variable in quantum CFD and revisited iteratively throughout the design pipeline, as different algorithmic components interact and influence one another.    
\end{abstract}

\section{Introduction}
Quantum computing aims to harness quantum mechanical phenomena, such as superposition and entanglement, to achieve computational capabilities beyond those of classical computers for certain tasks. Since its inception over four decades ago, the field of quantum computing has expanded rapidly, branching into a diverse array of disciplines beyond just simulating inherently quantum systems, as first proposed by Feynman \cite{feynman1982simulating}. While simulating many-body quantum systems remains a promising area of research for quantum computing \cite{fauseweh2024quantum}, additional applications are increasingly being explored across a wide range of classical domains \cite{bova2021commercial,hassija2020forthcoming,bayerstadler2021industry}, including materials and pharmaceuticals, automotive manufacturing, satellite communication and risk analysis. Much of the enthusiasm surrounding the field has been driven by improvements in quantum hardware but also seminal quantum algorithms, with notable examples such as Grover's algorithm \cite{grover1997quantum} for search and Shor's algorithm \cite{shor1994algorithms} for factoring, which collectively sparked widespread excitement by demonstrating the potential for quantum advantage over their classical counterparts.

Similarly, computational fluid dynamics (CFD) plays a critical role across a wide range of applications with profound practical importance \cite{blazek2015computational,sharma2011review,zawawi2018review}, including aerodynamics, weather prediction, biotechnology, and industrial design. While CFD has significantly benefited from decades of advances in classical computing, especially through increases in processor clock speeds and parallel architectures, the slowing of Moore's Law has shifted attention toward algorithmic innovation and the development of alternative computational accelerators \cite{elmisaoui2023high,wang2024recent} with quantum computing being a key candidate \cite{malinverno2025review,amaral2025quantum}. This shift is particularly crucial given the enormous computational demands inherent to practical CFD problems.

Take, for example, the simulation of turbulent flows using Direct Numerical Simulation (DNS), where the computational complexity scales roughly as $Re^3$, with $Re$ denoting the Reynolds number. In realistic configurations, such as the simulation of airflow around an aircraft, $Re$ can have values on the order of $10^8$, suggesting a requirement of approximately $10^{24}$ operations --- pushing the limits of even an ideal exascale classical supercomputer \cite{succi2023quantum}. Meteorological and climate simulations involve even larger Reynolds numbers by several orders of magnitude \cite{sprague2017turbulent}, placing them well beyond the realm of classical DNS.

To appreciate the potential role quantum computing \emph{might} have in addressing these challenges, consider the unrealistically optimistic perspective of how this $Re^3$ classical complexity could be compressed using qubits rather than bits. Assuming that the flow state can be efficiently encoded into the amplitudes of a quantum state, the number of qubits ($Q$) required to represent the system scales logarithmically
\begin{equation}
    Q = 3 \log_2 Re.
\end{equation}
This suggests the potential for a dramatic reduction in storage complexity. For instance simulating an aircraft-scale flow with $Re \approx 10^8$ might, utilizing an idealized quantum encoding, be represented using just $\sim 80$ fault tolerant qubits \cite{succi2023quantum}. However, achieving a quantum advantage in practice is subject to massive caveats, the choice of encoding in particular is usually of paramount importance, often resulting simultaneously in the algorithm's greatest strength and weakness. 

The aim of this work is to provide a high-level assessment of how information can be encoded for fluid simulations intended to run on quantum hardware, and to clarify where the key opportunities and challenges lie.
Rather than committing to a single algorithm or full end-to-end implementation, we adopt an architecture agnostic perspective that allows comparing multiple encoding strategies while preserving the broader picture, which would be impractical if each option were implemented in detail. 
Our objective is to identify promising directions and discuss bottlenecks, so that subsequent, focused research can target the most impactful issues.

To keep the scope of this work manageable, we will only consider direct simulation of fluids evolving over time using the quantum computer, rather than cases where a quantum computer is used to solve an important subproblem as was done in \cite{rathore2025integrating,rathore2025load}. Although the use of quantum subroutines is a promising avenue for the application of quantum computing \cite{callison2022hybrid}, there are too many possibilities to consider and they will involve information being processed in very different ways. 

The remainder of this work is organized as follows. Section \ref{A:1} discusses the requirements that CFD simulations impose and how these relate to quantum computing, where we keep the discussion deliberately broad to encompass as many classical methods as possible. Section \ref{A:2} then reviews several widely used state encoding strategies. Next, Section \ref{A:3} presents a set of quantum algorithms for fluid simulation, each illustrating a distinct facet of quantum encoding. For example, Section \ref{sec:1} addresses the measurement bottleneck inherent to amplitude encoding, Section \ref{sec:2} outlines approaches for handling nonlinearity, Section \ref{sec:3} showcases the use of basis encodings, and Section \ref{sec:4} examines cases where the optimal encoding is not obvious. Finally, Section \ref{A:4} offers a summary and an outlook on future directions. 

\section{Quantum-Enhanced Fluid Simulations: A Method-Agnostic Perspective}\label{A:1}
In this section, we outline the essential requirements that classical fluid dynamics simulations impose on quantum algorithms. To maintain generality, we deliberately avoid focusing on method-specific frameworks and instead highlight the broad expectations shared by the CFD community when simulating fluid behavior.

\subsection{Requirements For Quantum Representation Of Fluids}\label{s22}
A fundamental challenge in applying quantum computing to CFD lies in bridging the classical and quantum realms. This primarily involves two critical processes, namely loading classical data into quantum states (state preparation) and extracting meaningful classical information from quantum states (state measurement). 

CFD inherently involves modeling the evolution of field quantities such as velocity components, density, and energy across a spatial domain. In more complex scenarios, such as reactive or multi-physics flows, this can extend to hundreds of variables representing chemical species, temperature, or other physical quantities. While initializing these values on classical hardware is straightforward to the point of trivial, preparing a quantum state that coherently encodes the same classical data distribution can be very challenging. 
The complexity of a quantum circuit is often quantified in terms of the number of two-qubit operations, particularly CNOT gates. The reason for this is that these gates are experimentally expensive to implement and susceptible to noise, making them a limiting factor in quantum hardware performance \cite{murairi2022many}. Given that preparing an arbitrary quantum state can require an exponential number of CNOTs \cite{plesch2011quantum} and in the worst case nullify any computational advantage offered by the quantum method itself, various strategies have been proposed to mitigate this cost \cite{aulicino2022state,araujo2021divide,sanders2019black}. However, extending such approaches, particularly for general purpose state preparation, remains an open challenge.

In addition to gate count, circuit depth, i.e. the number of sequential layers of quantum operations, is another critical constraint \cite{bhattacharjee2019muqut}. On noisy intermediate scale quantum (NISQ) devices, deeper circuits are more prone to decoherence and error accumulation. Without auxiliary (ancilla) qubits, the depth required to prepare a general quantum state is inevitably exponential \cite{zhang2022quantum}. Introducing ancilla qubits can reduce depth to subexponential scaling, but this improvement comes at the cost of increasing circuit width (i.e. number of qubits used), which in the worst case can also scale exponentially. 
Optimizing this trade-off between time and space complexity can be important, especially for fluid simulations that demand high-dimensional state representations.
Accordingly, the cost of state preparation deserves explicit consideration in analyses of quantum algorithms. In many cases, complexity estimates assume that the initial quantum state can be prepared efficiently or is available from an earlier quantum routine. For CFD problems, however, this state will often need to encode classical initial conditions directly, making preparation costs an integral part of the algorithm. Rather than being secondary implementation details, these costs follow directly from the selected encoding strategy and should form part of any realistic performance assessment.

A second major challenge in applying quantum computing to CFD is the task of extracting meaningful classical information from a quantum state. The final outputs of interest in CFD, such as the fluid's velocity or pressure fields, are inherently classical. Therefore, a quantum state that encodes such data is only useful if it can be reliably measured and interpreted. Since quantum measurement generally alters or collapses the state, a large number of identical copies of the state are required to extract statistical information with high confidence.
The standard procedure for fully reconstructing a quantum state is known as \emph{Quantum State Tomography (QST)}. QST involves performing an informationally complete set of measurements that together can uniquely determine the state. Despite being theoretically robust, this process is resource intensive, with a cost that scales exponentially with the number of qubits \cite{paris2004quantum}. This exponential scaling makes full-state tomography impractical for all but the smallest quantum systems. This highlights a critical trade-off: while certain encodings may reduce the number of qubits needed to represent data, they might also increase the cost and complexity of extracting useful classical quantities. 

Most CFD applications also require time stepping to capture the temporal evolution of the system, introducing an additional layer of complexity for quantum algorithms. In many cases, this implies that quantum state preparation and measurement may be required at each time step, potentially increasing the computational overhead significantly.
One possible strategy to avoid this is to focus solely on solutions, which do not require detailed information at intermediate time steps and instead only measure at the end of the simulation. However, this approach is somewhat limited in scope. First, the most scientifically and practically relevant CFD problems exhibit rich transient dynamics, and accurately simulating these dynamics is frequently the primary goal. Second, many CFD problems involve coupled systems of partial differential equations, where the evolution of one variable depends on the time-dependent behavior of others. These dependencies impose a sequential structure that necessitates access to intermediate solutions at each time step, further increasing the depth and complexity of any quantum circuit designed to simulate full temporal evolution, unless the entire problem can be encoded efficiently.
As a result, the simulation of time-dependent CFD problems on quantum hardware will likely require new algorithmic and encoding/decoding strategies to manage the cost of repeated state preparation and measurement without compromising the fidelity or scalability of the computation. 
 
Boundary conditions are a fundamental component of any meaningful CFD simulation, as evidenced by the extensive body of literature dedicated to their development and implementation in classical CFD \cite{zawawi2018review,rempfer2006boundary,mittal2005immersed}. These conditions can range in complexity from inflow or outflow specifications to the incorporation of solid geometries, which may themselves be dynamic and time-dependent.
In the current landscape of quantum algorithms for CFD, still in a relatively nascent stage, the treatment of boundary conditions is often simplified or neglected. Many quantum approaches default to periodic boundary conditions, which, while mathematically convenient and applicable in some contexts, do not capture the diversity of boundary phenomena encountered in practical applications.
Looking ahead to real-world implementation, it is clear that quantum CFD solvers will require versatile and robust boundary treatments, capable of addressing non-periodic, complex, and perhaps even non-static boundaries. Furthermore, the choice of quantum encoding can significantly influence the difficulty of implementing boundary conditions. Certain encodings may reduce structural constraints or overhead when implementing boundary treatments, as discussed in subsequent sections, yet this caveat is often not explored when resorting to only periodic boundaries. 

Although CFD encompasses a wide range of formulations depending on the physical regime and application, it is illustrative to consider the Navier-Stokes equations as a representative system of governing equations due to their generality. They are commonly expressed as,
\begin{equation}
    \frac{\partial \rho}{\partial t} + \nabla \cdot (\rho\, \mathbf{u}) = 0,
\end{equation}
\begin{equation}\label{e:1}
    \frac{\partial \mathbf{u}}{\partial t} + (\mathbf{u} \cdot \nabla) \mathbf{u} = -\frac{1}{\rho}\nabla p + \nu \nabla^2 \mathbf{u} + \mathbf{f},
\end{equation}
where $\mathbf{u}$ denotes the velocity field, $p$ is the pressure, $\nu$ is the kinematic viscosity, and $\mathbf{f}$ represents external body forces. 
These equations highlight two further challenges that arise when attempting to simulate classical fluids using quantum algorithms, namely \emph{nonlinearity} and \emph{dissipation}. The convective term $((\mathbf{u} \cdot \nabla) \mathbf{u})$ introduces nonlinearity, while the viscous diffusion term ($\nu \nabla^2 \mathbf{u}$) introduces irreversibility and energy dissipation.
Quantum mechanics, in contrast, is governed by unitary evolution under the Schr\"odinger equation. This evolution is linear and norm-preserving, and is defined via a Hermitian Hamiltonian operator \cite{nielsen2010quantum}. As such, reconciling the nonlinear and dissipative nature of the Navier-Stokes equations with the linear, reversible dynamics of quantum computing presents a significant challenge.

The issue of dissipation, although nontrivial, can be conceivably addressed using several possible frameworks. For example, a potential strategy is to embed the non-Hermitian operator associated with dissipation into a higher-dimensional Hermitian form, by constructing a block-Hermitian matrix that includes both the operator and its Hermitian conjugate. This allows the use of standard Hamiltonian simulation techniques and preserves the essential dissipative behavior in an indirect fashion. 

Meanwhile, nonlinearity presents a more fundamental challenge for quantum computing. In classical computing, nonlinear terms such as $u^2$, which arise naturally in fluid dynamics (e.g. from the convective term in Equation~\ref{e:1}), can be computed with ease. This typically involves creating a temporary copy of a variable, multiplying it by the original, and storing the result. Such operations are trivial and ubiquitous in numerical solvers.
In quantum computing, this approach is fundamentally prohibited by the \emph{no-cloning theorem}, which states that it is impossible to create an identical copy of an arbitrary unknown quantum state \cite{wootters1982single}. As a result, operations that rely on duplicated data, such as those required for direct nonlinear term evaluation, are not immediately available on quantum hardware.
Even when indirect methods such as nonunitary evolution via measurements or open system dynamics  are possible, the feasibility of modeling nonlinearity is heavily influenced by the choice of encoding. Encoding choices not only determine the efficiency of quantum operations, but also the accessibility of nonlinear transformations. For example, encoding strategies that are optimal for linear operations (e.g. amplitude encoding) often become cumbersome or intractable when extended to nonlinear dynamics. Conversely, more expressive encodings that can support limited forms of nonlinearity may require greater quantum resources. 
Strategies to address nonlinearity in quantum algorithms are currently under active investigation \cite{kyriienko2021solving,tennie2025quantum,costa2025further}. However, it is important to recognize that sometimes these methods are inherently constrained not only by the linear, unitary foundations of quantum mechanics but also by the structural limitations imposed by the encoding paradigm.

\subsection{Defining a Quantum Advantage}

A central goal in the quantum computing community is to demonstrate that quantum computers can solve problems that are fundamentally intractable for classical machines. This pursuit has led to notable achievements \cite{boixo2018characterizing,arute2019quantum}, though many open questions remain \cite{markov2018quantum,harrow2017quantum}. A limitation of many of these demonstrations is that there is no requirement for the problems being solved to be practically useful.
In contrast, the CFD community is intrinsically focused on solving physically meaningful problems.
Since the best possible classical algorithm is often unknown in this domain, quantum advantage is more realistically defined relative to the best currently known classical methods, while acknowledging that this baseline may improve over time.

Within the landscape of quantum speedup and complexity theory, various theoretical frameworks exist to evaluate algorithmic improvements, whether in terms of computational resources, time to solution, or other metrics. The \emph{oracle model} is a well known example (see Supplementary Information for more background details), with Grover's algorithm offering a provable quadratic speedup over any classical alternative for unstructured search problems in the number of oracle calls. Yet, in practical applications like CFD, the concept of an oracle can lack a clear or meaningful physical interpretation, complicating its implementation. 
Similarly quantum complexity analysis often makes a series of assumptions about the nature of the problem that can create a gap between theoretical and realistically attainable advantages.
This work does not delve into the formal intricacies of complexity theory. Instead, we focus on the practical challenges of translating theoretical quantum advantages into real world performance gains for CFD, with particular attention given to how encoding strategies influence this transition.

As an illustrative example, consider the well-known quantum algorithm proposed by Harrow, Hassidim, and Lloyd \cite{harrow2009quantum}, commonly referred to as HHL, for solving a linear system of equations (LSE), defined as:
\begin{equation}
    \mathbf{Ax} = \mathbf{b},
\end{equation}
where $\mathbf{A}$ is an $N \times N$ coefficient matrix, $\mathbf{b}$ is a known vector, and $\mathbf{x}$ is the solution vector.
The HHL algorithm achieves a runtime scaling of $\mathcal{O}(\log(N) s^2 \kappa^2 / \epsilon)$, where $\epsilon$ denotes the desired precision, $\kappa$ is the condition number of the matrix, and $s$ represents its sparsity. This implies an exponential speedup in terms of problem size relative to classical algorithms such as conjugate gradient methods~\cite{shewchuk1994introduction}. Given the ubiquitous role of LSEs across scientific and engineering applications, this result generated widespread interest in the community, leading to numerous proposed applications \cite{ye2024hybrid,lapworth2022hybrid,gopalakrishnan2024solving}.

Subsequent work has improved several aspects of HHL, including the dependence on precision \cite{childs2017quantum}, sparsity assumptions \cite{wossnig2018quantum}, and condition number scaling \cite{singh2024optimised,tsemo2024enhancing}. Nevertheless, a practical quantum speedup for CFD problems formulated directly as linear systems remains elusive. The main reason is that the formal complexity advantage depends on assumptions whose costs may dominate in realistic settings, especially the efficient preparation of the input state $\mathbf{b}$ and the extraction of useful information from the output state encoding $\mathbf{x}$. As emphasized in the previous subsection, both costs are strongly encoding-dependent.

Consequently, a naive formulation that simply maps the problem directly onto a linear system of equations, without careful consideration of encoding, is unlikely to yield a meaningful quantum speedup in practice, especially when full reconstruction of $\mathbf{x}$ is required. A more promising direction is to adopt an encoding-aware design in which the quantum representation is tailored to specific observables or statistics of interest, since these may be estimated more efficiently \cite{huang2020predicting,zhang2021experimental}. However, extending such restricted outputs to meaningfully support practically relevant CFD tasks remains an open challenge. Moreover, once the objective is relaxed from solving the full linear system to estimating selected observables, it is not always clear that a quantum method retains an advantage over the best classical alternatives \cite{aaronson2015read}, making it important to ensure that any comparison is made on a fair basis.
On the input side, many CFD problems also do not naturally guarantee \emph{a priori} that the vector $\mathbf{b}$ has structure amenable to efficient quantum state preparation. Taken together, these observations do not necessarily rule out HHL-inspired approaches, but rather emphasise the importance of encoding as a central design principle. From this perspective, a versatile approach can be to focus on targeted subproblems instead of the full system, particularly where the quantity of interest can be accessed without reconstructing the full classical solution \cite{rathore2025integrating}.


While perhaps not directly applicable to CFD, it is worth noting that Shor's algorithm \cite{shor1994algorithms} serves as a compelling example of how quantum algorithms can efficiently navigate these input/output bottlenecks. The input state is a simple uniform superposition, which is straightforward to prepare, and the quantum computation produces an output from which only a small amount of classical information, specifically an estimate of a periodicity parameter, needs to be extracted. This is possible because the goal is to find an integer $r$, close to $qc/r$, where $q$ is the size of a Hilbert space. The output state is a superposition of all possible values of $c$, but by construction a measurement collapses this into a single random integer which allows determining $r$ with high probability through the use of a continued fraction expansion \emph{a posteriori}.
This limited quantum-classical communication is sufficient for a complete classical reconstruction of the solution \emph{a posteriori}, thereby preserving the exponential quantum speedup.

These examples illustrate that, in CFD, a meaningful quantum advantage cannot be assessed from simplified asymptotic runtime analysis of the subroutine alone. It is also essential to account for the hidden encoding-dependent costs such as state preparation, intermediate access during time evolution, and final measurement. These overheads are not merely technical caveats: they can determine whether a theoretical speedup survives in practice. Different encodings therefore lead to different trade-offs among efficiency, flexibility, and compatibility with problem-specific requirements such as boundary treatment or nonlinear dynamics, themes that will be developed in the following sections.


\section{Theoretical Background For Encoding Strategies}\label{A:2}
This section introduces the principal encoding schemes that underpin the applications discussed in later sections. In particular, amplitude encoding forms the basis of the methods examined in Sections \ref{sec:1} and \ref{sec:2}, whereas Section \ref{sec:3} illustrates the potential of basis encodings. Section \ref{sec:4} incorporates aspects of both, while also making use of block encodings. The subsequent discussion of temporal encoding is, in general, compatible with most methods.
\subsection{Amplitude Encoding}
Amplitude encoding is perhaps the most commonly employed choice in quantum enhanced CFD applications \cite{budinski2021quantum,bharadwaj2023hybrid}. Its popularity stems largely from the ability of a quantum register employing amplitude encoding to efficiently leverage the tensor product structure such that the the number of possible configurations scale exponentially with register size, resulting in the ability to explore a state space that grows exponentially faster than would be possible classically. 

In amplitude encoding, a vector of real numbers,  
\begin{equation}
    \vec{X} = (x_0, x_1, \dots, x_{N-1}), \quad x_k \in [0,1]
\end{equation}
is mapped to the amplitudes of a quantum state:  
\begin{equation}
\vec{X} \rightarrow \sum_{k=0}^{N-1} x_k \ket{k}, 
\end{equation}
where a separate normalization factor ensures the validity of the quantum state. This approach allows an \( n \)-qubit register to store up to $2^n$ complex numbers simultaneously within a superposition state. Its appeal lies not only in this compact storage, but also in its compatibility with several prominent quantum algorithmic primitives, particularly those based on linear algebra and unitary evolution. For these reasons, amplitude encoding often provides the clearest route to formal quantum speedups. At the same time, realizing these advantages in practical CFD settings faces substantial challenges related to input, output, and nonlinear dynamics.

As discussed in Section~\ref{s22}, extracting full classical information from a generic quantum state is generally expensive and may require quantum state tomography \cite{cramer2010efficient}. This issue is especially acute for amplitude encoding, since compressing $2^n$ coefficients into a $n$-qubit state does not necessarily imply that those coefficients can be recovered efficiently. Indeed, many amplitudes in a high-dimensional state may be exponentially small, so resolving them with high confidence can require an exponential number of repeated measurements and state preparations. 
Even after which, the computationally demanding inverse reconstruction problem must also be solved classically. Although exploitable structure may sometimes reduce the tomography overhead \cite{holmes2020efficient,zylberman2024efficient}, this is not always viable in general, but is currently an area of active research. Similarly, while selected statistics or observables may be estimated efficiently \cite{huang2020predicting,zhang2021experimental,jin2022quantum,torlai2020precise}, determining which such quantities are sufficient for practically relevant CFD tasks remains an open question. Thus, for amplitude encoding, the central challenge is not only how compactly the data can be represented, but also whether the necessary post-processing can be performed without eroding any quantum advantage. How efficiently this can be done is a consequence of both the structure of the data but also the chosen encoding itself.

Preparing a quantum state with amplitude encoded classical data is also a nontrivial task.  Although several methods have been proposed \cite{jaques2023qram,gonzalez2024efficient}, they often comes at the cost of an exponential number of resources \cite{mottonen2004transformation,sun2023asymptotically}, reliance on oracles \cite{grover2000synthesis,bausch2022fast}, or assumptions regarding inherent structure in the input data \cite{zhang2022quantum,holmes2020efficient}. For unstructured data it is frequently \emph{assumed} that the desired state can be prepared \emph{efficiently} through the use of Quantum Random Access Memory (QRAM). QRAM serves as the quantum analogue to classical  RAM, enabling access to memory addresses in coherent superposition. While QRAM could in theory, efficiently prepare an amplitude encoded state in logarithmic time, the physical resources required scale exponentially with respect to the number of qubits \cite{ciliberto2018quantum}. Consequently, the practical feasibility of QRAM remains an contentious issue in the quantum computing community. Moreover, even if such a device could be built, it is unclear whether a parallel, classical system given the same resources could outperform its QRAM counterpart \cite{Steiger2016RacingIP,aaronson2015read}. Therefore, assumptions of efficient amplitude encoding in the absence of well structured data or alternative mechanisms for efficient state preparation should be treated with care.

A key challenge seen in amplitude encoding for CFD is how to engineer interactions between different parts of the wavefunction. Since quantum mechanics is a fundamentally linear theory, implementing the kind of nonlinear interactions needed in fluid mechanics necessitates advanced methods, using measurement and post-selection for example. 
Although it seems natural that an efficient encoding should always be preferable, it is generally difficult to implement arbitrary interactions, and particularly nonlinear interactions between quantum amplitudes in different states. 
Adding even a small degree of non-linearity to fundamental quantum operations would imply that quantum computers could solve NP-complete problems in polynomial time \cite{Abrams1998nonlinearNP}. If we assume that the widely believed conjecture that quantum computers are not capable of solving NP-complete problems in polynomial time, then this implies that nonlinearity between amplitudes within a quantum register cannot be easily simulated and implementing non-linearity within a single register would have to have sufficiently poor scaling to introduce an exponential overhead to the algorithm in \cite{Abrams1998nonlinearNP}. Alternative possibilities include a simulation that scales well but is unable to attain a speedup due to excessive error accumulation, or forms of nonlinearity where the arguments of \cite{Abrams1998nonlinearNP} do not hold. Otherwise a quantum computer based on ordinary (linear) quantum mechanics could simply simulate a nonlinear device, and  by using this simulation method NP-complete problems could be solved in polynomial time.

These observations highlight that the usefulness of amplitude encoding for fluid simulations depends strongly on the structure of the target problem.
In particular, while compact state representation is an important advantage, it does not by itself guarantee that the interactions required by nonlinear fluid dynamics can be implemented efficiently. Amplitude encoding may therefore be especially attractive in settings dominated by linear structure, or in workflows where only limited derived information must be extracted, whereas its role in fully nonlinear CFD remains a more open and problem-dependent question.

\subsection{Basis Encoding}
Basis encoding is the most straightforward approach, in which each input is mapped directly onto a computational basis state of the quantum register. The specific mapping can take several forms \cite{dominguez2023encoding,tamura2021performance}. One simple example is \emph{unary encoding}, in which a numerical value is represented by the number of qubits occupying a given state. For instance, the number $2$ may be represented by $\ket{1100}$, while the number $3$ may be represented by $\ket{1110}$, with the value determined by the number of qubits in the $\ket{1}$ state. This is clearly less space efficient than amplitude encoding, since the number of valid configurations in a unary register grows only linearly with the size of the register.

Despite this reduced compactness, unary encodings can still be very useful in practice, particularly for analog quantum machines. The reason is that they allow arbitrary pairwise interactions between logical variables to be expressed using only quadratic terms in the problem Hamiltonian \cite{kendon2026quantum}. By contrast, such interactions cannot in general be reproduced using only linear and quadratic terms when a denser encoding is employed, making the trade-off between compactness and interaction simplicity an important one that has been studied extensively in \cite{chancellor2019domain,chen2021performance,berwald2023understanding}. In a gate-model setting, this could be interpreted as a trade-off between qubit count and circuit depth, since a larger register can sometimes simplify the structure of the required operations. A single unary register does not by itself provide a scalable model of quantum computation, but a system composed of a growing number of such registers can still possess the robust tensor product structure required for scalability \cite{Blume-Kohout2002scalable}.

A particularly useful special case of unary encoding is the \emph{one-hot encoding} \cite{hadfield2019quantum}, which has long been used in areas such as linear and combinatorial optimisation. In a one-hot encoding, a logical variable with $n$ possible values is represented using $n$ spins or qubits, together with quadratic constraints that enforce the condition that exactly one of them takes the value 1. Because each logical state is associated with a distinct qubit in the $\ket{1}$ state, arbitrary pairwise interactions between one-hot encoded variables can be implemented using only two body Ising couplings. This makes one-hot encoding especially natural for spin-based optimisation settings, where restricting the problem Hamiltonian to quadratic terms is often desirable.

More common, particularly in the context of QCFD, is the use of a binary encoding in the basis states. The classical data is first converted to binary strings and the string is then encoded into a set of qubits. For example the number $5$ can be expressed in binary as $101$ which corresponds to the state $\ket{101}$ of a three qubit register. More generally, the number $x_i$ can be expressed as the bit string $\sum_{i=k}^nb_i2^i$ which can then be encoded into the state $\ket{b_n...b_k}$. This is more efficient than a unary encoding as the number of possible valid states scales exponentially with register size. However, unlike amplitude encoding each bit of classical information requires one qubit. The number of required qubits to encode information is therefore directly dependent on the data itself, both in terms of numerical magnitude but also the desired precision.  

Preparing a single encoded basis state is easily achieved by applying bit flips to the relevant qubits, in essence requiring just one single parallel operation. Translating an entire dataset into basis encoding can be more challenging. Although it is possible to store information over a superposition of basis states, if the desired superposition is not uniform or desired over a custom subset of states, this would require controlling the amplitudes as well which then results in added state preparation cost. Despite this, a key advantage of encoding in basis states as opposed to amplitudes is the greater ease in engineering arithmetic operations between the encoded data.  

Note that for the majority of cases, CFD will involve the manipulation of real numbers and not just integers. Even though classical computers rely on encoding numbers in the floating-point format using the IEEE 754 international standard \cite{zuras2008ieee}, a similar standard for quantum computing does not yet exist despite being desirable. 
One possible approach is taking a direct analogy with classical representation of a floating-point number and encoding the mantissa as well as exponent into a basis state \cite{steijl2022quantum}. 
Subsequently, it is possible to define circuits for employing arithmetic based on the quantum Fourier transform, including evaluation of the nonlinear product terms commonly arising in CFD. 
While much of the previous work on quantum arithmetic has been focused on fixed point representations \cite{bhaskar2015quantum}, due to cheaper circuit implementations, the range of numbers that can be represented by a fixed set of qubits in this way remains limited. 
Floating point representations significantly extend this range and even though the resource requirements of floating point addition are much larger than for fixed point numbers, the cost of multiplication is roughly similar for both. Moreover, given that multiplication in both representations is more costly than addition, it is reasonable to consider multiplication as the measure of choice, suggesting that floating point representation is a viable candidate for quantum arithmetic \cite{haener2018quantum}. 

\subsection{Converting between encodings}\label{s:0}
\begin{figure*}[h!]
\centering
\includegraphics[width=0.8\textwidth,trim=1cm 3cm 3cm 2cm,clip]{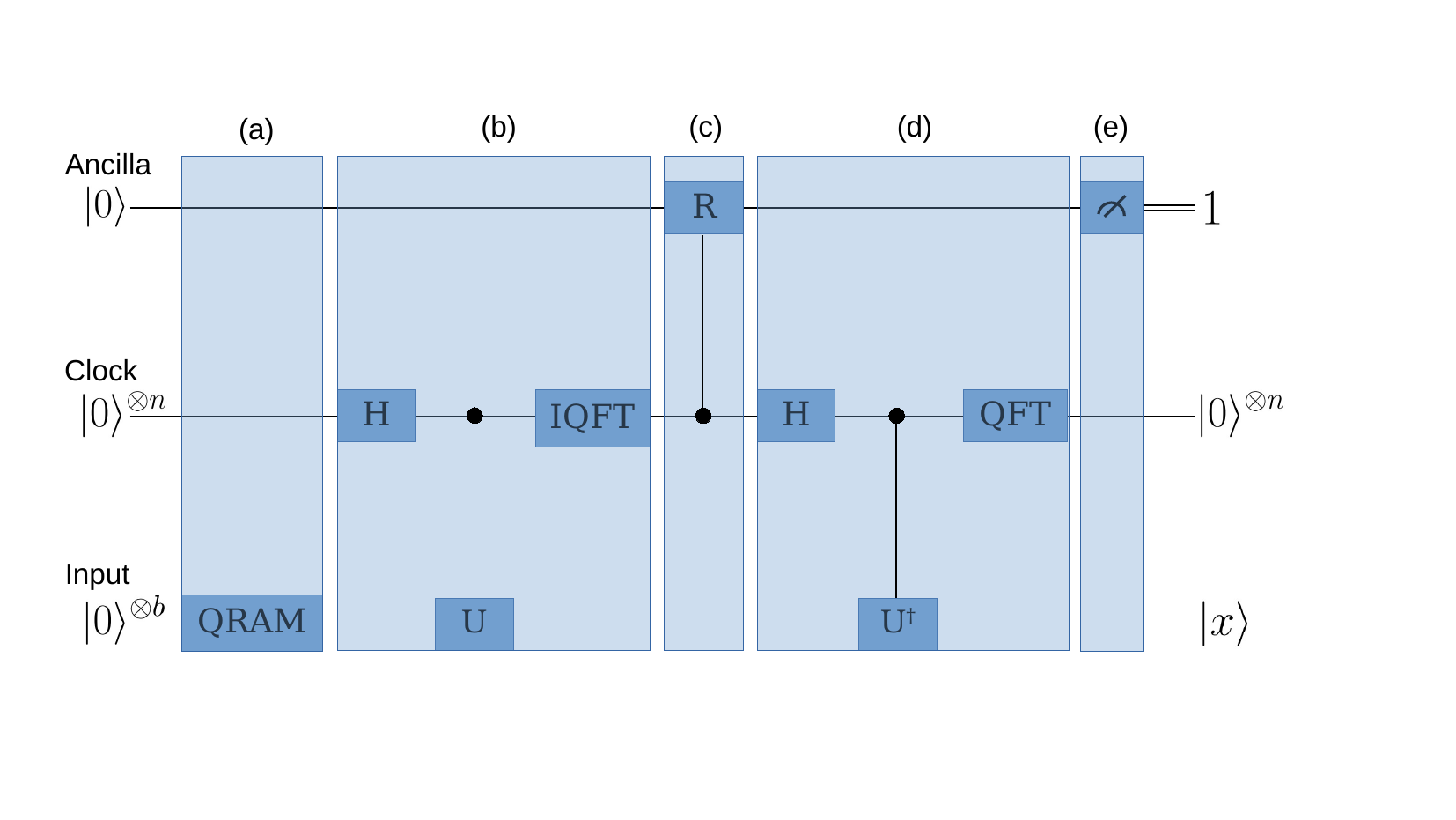}
\caption{Summary circuit diagram for the HHL algorithm, adapted from \cite{rathore2025integrating} (\href{https://creativecommons.org/licenses/by/4.0/}{CC BY}). The five main components are (a) State preparation, (b) Quantum phase estimation, (c) Controlled rotation, (d) Inverse quantum phase estimation and (e) Ancilla qubit measurement. Quantum phase estimation consists of Hadamard gates (H), controlled unitaries (U) and the inverse quantum Fourier transform (IQFT). Meanwhile, the input consists of three registers, referred to here as the ancilla, clock and input registers.}
\label{fig:1}
\end{figure*}

It can be useful to convert information encoded in the amplitudes of a quantum register into the basis states of another register, or vice versa. This transformation is not always obvious and can sometimes occur implicitly as part of a subroutine within a larger quantum algorithm. A canonical example is the HHL algorithm, illustrated in Figure~\ref{fig:1}.
The interested reader is referred to several pedagogical sources \cite{zaman2023step,liu2022survey} for a detailed, step by step walkthrough of the HHL circuit. 
Here we simply emphasize the movement of eigenvalues between different encodings. 
Specifically, QPE encodes eigenvalues of a matrix into the basis states of a dedicated clock (or phase) register. Thus, after applying QPE the system is in the state:
\begin{equation}
\Psi_b = \ket{b}^{\otimes b} \ket{\Tilde{\lambda}}^{\otimes n} \ket{0},
\end{equation}
where $\Tilde{\lambda}$ denotes an approximation of the scaled eigenvalues of the linear system problem matrix.

Next, a controlled rotation is applied to an ancillary qubit, conditioned on the clock register. This results in the state
\begin{equation}
\Psi_c = \ket{b_i}^{\otimes b} \ket{\Tilde{\lambda_i}}^{\otimes n} \left(
\sqrt{1 - \frac{C^2}{\Tilde{\lambda}_i^2}} \ket{0} + \frac{C}{\Tilde{\lambda}_i} \ket{1}
\right),
\end{equation}
where $C$ is a normalization constant. In this step, the scaled eigenvalues, which were originally encoded in the basis states of the clock register, are effectively transferred into the amplitudes of the ancillary qubit (as their inverse) and facilitate post-selection further downstream of the $\ket{1}$ state. An inverse QPE uncomputes the clock register, leaving the desired quantity encoded in the amplitudes of the system register. Together, these steps represent a mechanism for shuttling information between basis and amplitude encodings within a single algorithmic pipeline.

This approach can be generalized and results in a probabilistic algorithm for converting from a basis to amplitude encoding \cite{mitarai2019quantum}. Starting from a basis encoded state
\begin{equation}
    \frac{1}{\sqrt{N}}\sum_{j=1}^N\ket{j}\ket{\mathbf{d}_j},
    \end{equation}
where $\mathbf{d}_j$ is the binary encoded data, the first step is to calculate the angle $\psi_j=2/\pi\cos^{-1}d_j$ using quantum arithmetic \cite{ruiz2017quantum} such that
\begin{equation}
    \frac{1}{\sqrt{N}}\sum_{j=1}^N\ket{j}\ket{\mathbf{d}_j}\ket{0} \rightarrow \frac{1}{\sqrt{N}}\sum_{j=1}^N\ket{j}\ket{\mathbf{d}_j}\ket{\boldsymbol{\psi}_j} .
\end{equation}
Adding an ancilla and performing a controlled rotation $R_y=e^{i\pi \psi_j Y/2}$ results in the state
\begin{equation}
\frac{1}{\sqrt{N}}\sum_{j=1}^N\ket{j}\ket{\mathbf{d}_j}\ket{\boldsymbol{\psi}_j}\ket{0}_a \rightarrow \frac{1}{\sqrt{N}}\sum_{j=1}^N\ket{j}\ket{\mathbf{d}_j}\ket{\boldsymbol{\psi}_j}(d_j\ket{0}_a + \sqrt{1-d_j^2}\ket{1}_a), 
\end{equation}
where evidently a measurement of the ancilla in the $\ket{0}$ state results in the desired encoding of $d_j$ in amplitudes, after which the remaining registers can be disentangled by uncomputing the relevant operations. Naturally the requirement of post-measurement selection makes the algorithm probabilistic, and although this can somewhat be mitigated by amplitude amplification the feasibility for CFD data warrants further investigation. 

Alternatively, a more complex approach is required to convert data from an amplitude encoding to a basis encoding. Conceptually, this can be achieved by coherently performing a SWAP test between a register initialized with computational basis states (representing address indices) and a register containing the amplitude encoded state without performing a measurement. This process maps information about the amplitude overlaps into a structured quantum phase across the system's state vector.
A composite unitary operator can then be constructed whose eigenvalues encode these phase relationships. Applying quantum phase estimation to this operator allows the extraction of amplitude information into dedicated basis encoded registers. However, because quantum amplitudes generally contain both real and imaginary components, a complete conversion necessitates extracting and encoding each of these separately, as well as the magnitude. This requirement may impose significant overhead in terms of qubit resources and circuit depth, posing a nontrivial challenge for practical implementations of QCFD.
For a detailed technical treatment of this procedure, the reader is referred to \cite{mitarai2019quantum}.

\subsection{Block Encoding}
It is often necessary to encode structured or ordered data as a matrix, rather than as individual data points. A standard technique for enabling quantum processing of such matrices is to \emph{block encode} the input matrix $A \in \mathbb{C}^{N \times N}$, which may not be unitary, into a higher dimensional Hilbert space. This is done by embedding $A$ as the leading principal submatrix of a larger unitary operator $U$, such that
\begin{equation}\label{e:2}
U = \begin{bmatrix}
A / \alpha & * \\
* & *
\end{bmatrix},
\end{equation}
where $\alpha \ge \|A\|$ is a normalization constant. The unspecified entries, denoted by $*$, are arbitrary up to the constraint that $U$ must remain unitary.

The original matrix $A$ can then be accessed via appropriate projectors. Let $P_1 = \ket{0}\bra{0} \otimes I$ and $P_2 = \ket{0}\bra{0} \otimes I$ be projectors that isolate the encoded subspace. Then the matrix $A$ is related to $U$ via
\begin{equation}
A = \alpha \cdot P_1 U P_2.
\end{equation}

However, it is not enough simply knowing that a matrix can be block-encoded in principle. For the encoding to be useful in practice, the unitary $U$ from Equation \ref{e:2} must be efficiently implementable as a quantum circuit composed of elementary gates. Constructing such a circuit generally increases the size of the ancillary space and may incur significant gate overhead. For instance, existing schemes for block encoding arbitrary dense matrices \cite{chakraborty2018power,clader2023quantum} exhibit an exponential scaling of $\mathcal{O}(2^n)$ T gates with respect to the matrix dimension. More efficient constructions are possible by exploiting structure in the input matrix \cite{gilyen2019quantum,sunderhauf2024block}, but these typically require the use of oracles, complicating the assessment of their actual implementation cost.

Block encoding is most commonly used to prepare input matrices for the Quantum Singular Value Transformation (QSVT), a technique that allows polynomial transformations of a matrix's singular values. QSVT has led to a significant unification of quantum algorithmic frameworks, with many existing algorithms-such as HHL, Grover search, and quantum walks-being reformulated within this paradigm. While sometimes referred to as a ``grand unification of quantum algorithms'' \cite{gilyen2019quantum}, QSVT also underpins some recent applications in QCFD \cite{patterson2025measurement,lapworth2025preconditioned}. Nevertheless, it is important to recognize that the assumption of an efficiently realizable block encoding is not always justified for general matrices, resulting in an important potential obstacle to obtaining meaningful speedups for realistic configurations. In addition to the QSVT, block encoding is also accompanied by an associated calculus for summation and multiplication between matrices \cite{gilyen2019quantum}, is an essential component of Hamiltonian simulation \cite{low2019hamiltonian} and also used in the linear combination of unitaries (LCU) approach.  

\subsection{Temporal Encoding}
As the goal in QCFD is simulating fluids evolving over time, a central question is how to represent and advance time within the quantum framework. 
The most direct approach is to make the time undergone by the simulated system correspond to real time over which the simulation runs. 
In simulating time evolution sequentially each quantum circuit execution corresponds to one time step in the simulation, using a consistent set of qubits and gates across steps. 
This method is conceptually straightforward and leverages circuit modularity. However, in many practical CFD scenarios, the need to reinitialize or reload the quantum state between time steps can become a significant overhead, potentially negating the intended quantum speedup. 
It may be possible to mitigate some of these obstacles, particularly for high order implicit schemes (such as Runge-Kutta methods), by framing time-stepping as an optimization problem amenable for quantum annealers \cite{zanger2021quantum} as discussed in Section \ref{sec:3}. 
While somewhat less efficient, minimising error over each timestep could allow small corrections backwards in time and potentially improve stability. However, dealing with nonlinearity, managing hardware requirements and verifiably demonstrating quantum speedup guarantees persist as significant challenges.   

An alternative strategy is to encode the time dimension directly into the quantum state representation \cite{wang2025simulating,diaz2023parallel,bharadwaj2025compact,bharadwaj2023hybrid}. 
For example, a series of "clock" qubits that correspond to discrete time steps in superposition can be entangled with the main register, creating a combined history state that simultaneously describes the fluid state at multiple time steps. 
The joint state evolving over $T$ steps can then be represented by $\frac{1}{\sqrt{T}}\sum_{t=0}^{T-1}|t\rangle_{\text{time}}\otimes|\psi(t)\rangle_{\text{fluid}}$, where $|\psi(t)\rangle$ is the fluid state at time $t$. 

This approach offers a form of temporal compression, potentially allowing exponential savings with regards to the number of operations required to capture the full time evolution. However, this comes at the cost of increased circuit complexity, as each time-dependent update must be iteratively implemented via controlled unitaries conditioned on the clock register. Such designs often result in deeper circuits and are vulnerable to cumulative gate errors, decoherence or diminishing success probabilities over long simulated time horizons.
To address these limitations, hybrid schemes could be employed. For instance, a finite window of time steps can be encoded into the quantum state at once, trading full temporal compression for reduced circuit depth and improved robustness. Such sliding-window encodings allow local access to recent history without simulating the entire time trajectory in a single circuit execution.
In some applications such as linear system solvers, it may even be feasible to construct the full history state in one shot \cite{bharadwaj2023hybrid}, solving a global linear system that encodes the entire evolution. This method avoids step-by-step iteration and can mitigate the success probability decay typically associated with sequential approaches. While such one-shot encodings can improve algorithmic efficiency and stability, they often involve more elaborate circuit constructions, larger resource footprints and are not always viable for general algorithms.

Moreover, even though extracting certain global observables, such as mean quantities, can sometimes be done efficiently from a history state \cite{bharadwaj2023hybrid,bharadwaj2025compact}, recovering the full fluid configuration at a specific time slice typically still encounters the readout and resolution limitations intrinsic to the chosen spatial encoding. The choice of temporal encoding strategy thus involves trade-offs among circuit depth, qubit count, flexibility, and fidelity, all of which must be weighed in light of the target fluid problem and the capabilities of available quantum hardware.

\section{Quantum Enhanced Fluid Dynamics Simulation Methods}\label{A:3}

The discussion now turns to specific algorithms for fluid simulation, emphasizing how specific encoding choices shape their design and performance. The methods surveyed are not exhaustive, but instead are selected as archetypal cases, each of which showcase a different facet of encoding in QCFD. Section~\ref{sec:1} highlights one possible route to mitigating the measurement bottleneck associated with amplitude encoding and Section \ref{sec:2} examines approaches for incorporating nonlinearity. Section~\ref{sec:3} demonstrates the potential of basis encodings, while Section~\ref{sec:4} considers an algorithm in which no single encoding is clearly optimal, owing to conflicting requirements across its different algorithmic components.

\subsection{Solving Nonlinear PDEs Using Quantum Integration}\label{sec:1}
\begin{figure*}[h!]
    \centering
    \begin{subfigure}[t]{0.48\textwidth}
        \centering
        \includegraphics[width=\linewidth]{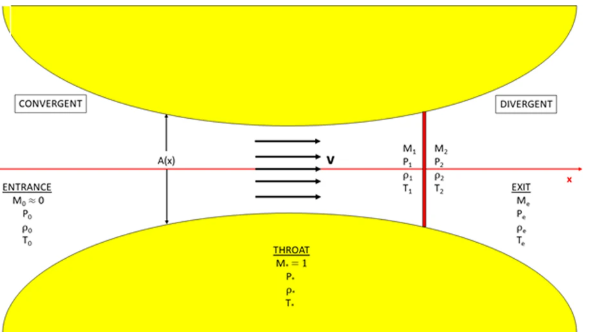}
        \caption{Schematic of flow through a convergent-divergent nozzle with a normal shock in the diverging section.}
    \end{subfigure}
    \hfill
    \begin{subfigure}[t]{0.48\textwidth}
        \centering
        \includegraphics[width=\linewidth,trim=9cm 5cm 10cm 3.7cm,clip]{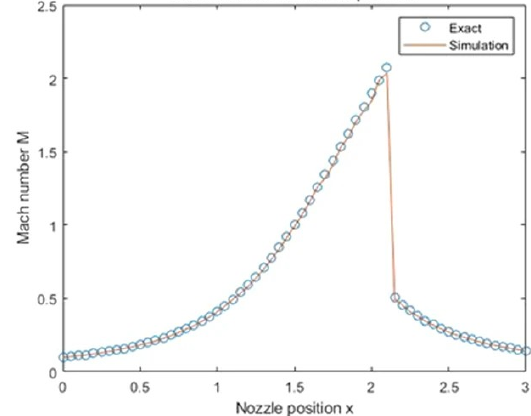}
        \caption{Mach number profile from the quantum Navier--Stokes simulation, compared with the exact solution.}
    \end{subfigure}

    \vspace{0.5em}

    \begin{subfigure}[t]{0.48\textwidth}
        \centering
        \includegraphics[width=\linewidth]{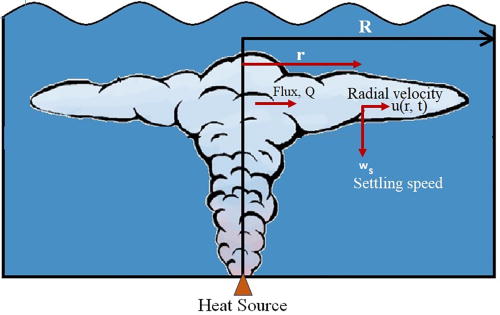}
        \caption{Schematic of tephra transport in a submarine volcanic eruption, showing the vertical stem and the horizontally buoyant umbrella region in which dispersal occurs.}
    \end{subfigure}
    \hfill
    \begin{subfigure}[t]{0.48\textwidth}
        \centering
        \includegraphics[width=\linewidth]{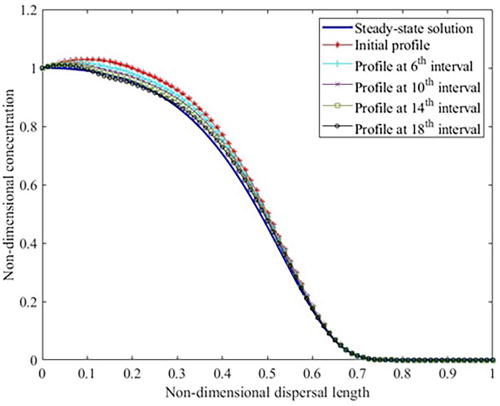}
        \caption{Spatial distribution of tephra concentration from the quantum-algorithm simulation, showing convergence toward the analytical steady-state solution.}
    \end{subfigure}
    
    \caption{Illustrative application examples of the Quantum Integration Algorithm to fluid and transport problems. Top: compressible flow through a convergent-divergent nozzle with a normal shock, adapted from \cite{gaitan2020finding}. Bottom: tephra transport in the umbrella region of a submarine volcanic eruption, adapted from \cite{basu2024quantum}. In each case, the left panel shows the physical setup and the right panel shows a representative numerical result. Figures licensed under \href{https://creativecommons.org/licenses/by/4.0/}{CC BY}.}
    \label{fig:2}
\end{figure*}

The quantum amplitude estimation (QAE) algorithm was introduced by Brassard \emph{et al.} \cite{brassard2002quantum} and demonstrated how quantum phase estimation (QPE) \cite{kitaev1995quantum} could be used to determine the average value of a Boolean function. The algorithm was subsequently extended \cite{novak2001quantum,heinrich2002quantum} to finding the average of a real-valued function, making it possible to approximate definite integrals. This quantum integration routine became the basis for a quantum ordinary differential equation (ODE) solver \cite{kacewicz2006almost}. However, it was only over a decade later that Gaitan \cite{gaitan2020finding} demonstrated the approach could be straightforwardly promoted to a PDE solver for a family of equations, including the Navier-Stokes. Subsequent studies have explored its application in a variety of configurations, several of which are illustrated in Figure~\ref{fig:2}. 

While a detailed explanation of the classical processing can be found in \cite{gaitan2020finding,kacewicz2006almost}, we simply outline these steps here before focusing on the quantum component of the algorithm. 
The quantum PDE solver seeks the approximate solution to a system of the form
\begin{equation}\label{e:3}
    \frac{\partial \mathbf{U}}{\partial t} = \mathbf{F}[\mathbf{U},\partial \mathbf{U}/ \partial x ... \partial^n \mathbf{U}/ \partial x_n],
\end{equation}
for a multi-component vector field $\mathbf{U}(\mathbf{x},t)$ over the time interval $0 \leq t \leq T$, with flux operator $\mathbf{F}$. 
The first step discretizes the system in space over $\mathbf{I}$ distinct grid points, as is commonly done in CFD, resulting in a reduction from a system of PDEs to the following system of ODEs, 
\begin{equation}\label{e:6}
        \frac{\mathrm{d}  \mathbf{U}(\mathbf{I},t)}{\mathrm{d} t} = \mathbf{f}[\mathbf{U}(\mathbf{I},t)].
\end{equation}
The driving map $\mathbf{f}$ is assumed to be a (generally nonlinear) H\"older class function of the state and its spatial derivatives. Choice of spatial scheme (e.g. finite difference/volume/element, order, stencil) alters only the form of $\mathbf{f}$ and not the subsequent quantum algorithm.

From here the quantum ODE solver developed previously by Kacewicz \cite{kacewicz2006almost} can be directly applied and consists of the following steps:  
\begin{itemize}
    \item Time Discretisation: The interval [0,T] is divided into n primary subintervals, each of which is further divided into $N_k$ secondary subintervals.
    \item Initialization: The algorithm initializes $y_o=U_o$ and introduces parameters $y_i \approx U(t_i)$ to approximate the solution at each primary subinterval boundary.
    \item Local Approximation: The solution is approximated using a truncated Taylor expansion within each secondary subinterval. 
    \item Solution Construction: The local Taylor approximations for each secondary subinterval are stitched together to build a continuous piecewise approximation, $\alpha (t)$, throughout each primary interval and consequently throughout the entire time interval once all the $y_i$ are known. 
    \item Integral Propagation: Each $y_i$ is determined iteratively in terms of $y_{i-1}$ and by approximating integrals of the form $\int f[\alpha(t)]dt$ over each primary subinterval, which in turn are broken down into sums of integrals over each of the secondary subintervals. 
\end{itemize}
The only quantum component of the algorithm is evaluating these integrals over the secondary intervals efficiently, with the remaining steps being purely classical. Note that the secondary subintervals are introduced so that the solution on each small time segment can be accurately approximated by a local Taylor expansion, while the update over the full primary interval can be rewritten as an average over many such local contributions. This is important because the latter average is the quantity estimated by the quantum amplitude estimation step, so the subdivision serves both an approximation role and a quantum-algorithmic one. This choice is also not arbitrary, Kacewicz \cite{kacewicz2006almost} showed that using such subintervals yields an almost optimal quantum algorithm.

Due to its inherent versatility, it is useful to first discuss amplitude estimation in a general setting prior to highlighting the role it plays in the quantum integration algorithm. Readers already familiar with the algorithm may choose to move forwards to Equation \ref{e:00} to see directly how this is incorporated into a PDE solver. 

Consider a quantum algorithm that prepares a state $\psi$ through some unitary transformation $A$ acting on an initial zero state. The Hilbert space ($\mathcal{H}$) can partitioned into the sum of a \emph{good} ($\mathcal{H}_1$) subspace and its orthogonal complement, a \emph{bad} ($\mathcal{H}_0$) subspace through a Boolean function ($\chi$), such that the $\ket{j}$ basis state is defined as good if $\chi (j) =1 $ and bad otherwise. Thus, the state $\ket{\psi}$ can be uniquely decomposed into its projections on these good and bad subspaces as 
\begin{equation}\label{e:99}
    A \ket{0}= \ket{\psi}= \sqrt{a}\ket{n_1}+\sqrt{1-a}\ket{n_0},
\end{equation}
where $\ket{n_1}$ ($\ket{n_0}$) represents the normalized component corresponding to to good (bad) outcomes and $\sqrt{a}$ is the probability amplitude associated with the good subspace, which we seek to estimate with QAE. This is a normalised quantum state and can be parameterized by an angle such that $\sin^2(\theta)=a$ to give the equivalent form 
\begin{equation}
    \ket{\psi}=\sin{\theta} \ket{n_1}+\cos{\theta}\ket{n_0} .
\end{equation}

The core mechanism of QAE is the Grover-like operator $Q$, which acts as a rotation in the two-dimensional subspace 
$\mathcal{H}_\psi = \text{span}\{ \ket{n_0},\ket{n_1} \}$. 
This rotation is implemented as a composition of two reflections
\begin{equation}\label{e:4}
Q = U_\psi U_{n_0},
\end{equation}
where $U_\psi = I - 2\ketbra{\psi}{\psi}$ is a reflection about the prepared state, and $U_{n_0} = I - 2\ketbra{n_0}{n_0}$ is a reflection about the normalized "bad" component of the state. 
Since the product of two reflections in a plane is a rotation, $Q$ acts within $\mathcal{H}_\psi$ as a rotation by angle $2\theta$,and is represented in the basis $\{ \ket{n_0}, \ket{n_1}\}$ as
\begin{equation}
Q =
\begin{pmatrix}
\cos(2\theta) & -\sin(2\theta) \\
\sin(2\theta) & \cos(2\theta)
\end{pmatrix}.
\end{equation}

The action of this operator in the full computational basis is more easily understood by considering its equivalent form
\begin{equation}\label{e:0}
    Q=-AS_0A^{-1}S_\chi,
\end{equation}
where $S_0$ flips the sign of the $\ket{0}$ state only and $S_\chi$ flips the sign of computational states labeled as good states only ($S_\chi=(-1)^{\chi{(j)}}\ket{j}$. Equation \ref{e:0} is equivalent to Equation \ref{e:4} as $AS_oA^{-1}=U_\psi$ and $-S_\chi=U_{n_o}$. 
However, this decomposition is useful for analyzing the action of $Q$ on the orthogonal complement $\mathcal{H}_\psi^\perp$. Since $AS_0A^{-1}$ acts as the identity on $\mathcal{H}_\psi^\perp$, the action of $Q$ in this subspace reduces to $-S_\chi$. Consequently, $Q^2 = I$ on $\mathcal{H}_\psi^\perp$, and all eigenvalues of $Q$ in this subspace are $\pm 1$. This implies that for an arbitrary initial state, which by construction has no initial overlap with $H_\psi^\perp$, it suffices to consider the evolution under $Q$ projected into the invariant two-dimensional subspace $\mathcal{H}_\psi$.

Thus, for the nontrivial case where $a\notin\{0,1\}$, $H_\psi$ is a two dimensional, \emph{stable} subspace with eigenvectors
\begin{equation}
    \psi_\pm=\frac{1}{\sqrt{2}}(\ket{n_1}\pm i\ket{n_0}),
\end{equation}
and eigenvalues
\begin{equation}
    \lambda_\pm=e^{\pm2i\theta}. 
\end{equation}
As these form a complete orthonormal basis for $H_\psi$, the initial state can be expressed in the eigenbasis by inverting these equations to give 
\begin{equation}
    \ket{\psi}=\frac{-i}{\sqrt{2}}(e^{i\theta}\ket{\psi_+}-e^{-i\theta}\ket{\psi_-}).
\end{equation}
Moreover, the effect of repeated applications of $Q$ is 
\begin{equation}\label{e:5}
    Q^j\ket{\psi}=\frac{-i}{\sqrt{2}}(e^{(2j+1)i\theta}\ket{\psi_+}-e^{-(2j+1)i\theta}\ket{\psi_-})=\sin[{(2j+1)\theta}]\ket{n_1} + \cos[{(2j+1)\theta}]\ket{n_0}.
\end{equation}
Equation \ref{e:5} illustrates the core principle of quantum amplitude amplification, where repeated applications of $Q$ can rotate an initial state such that the outcome $\ket{n_1}$ is obtained with high probability upon measurement, by choosing $j$ such that $(2j+1)\theta \approx \pi/2$ even if $\theta$ itself is not known. 
However, the goal of QAE is to estimate the amplitude which is now conveniently encoded in the phases of the eigenvalues of a unitary in Equation \ref{e:5}. Therefore, this can be efficiently extracted through a sighlty modified application of the QPE algorithm. 
Note that this assumes the use of an oracle whose action is defined as 
\begin{equation}
    \Lambda(Q)\ket{j}\otimes\ket{k}=\ket{j}\otimes Q^j\ket{k}.
\end{equation}
It is also convenient to define the Fourier state $\ket{S(\omega)}=\frac{1}{\sqrt{N}}\sum_{k=0}^{N-1}e^{2\pi i \omega k} \ket{k}$ such that $\ket{S(j/N)}=\text{QFT}\ket{j}$ for any $j \in \{0,1 \dots N-1{}\}$. The steps for QAE are then as follows

\begin{algorithm}[h!]
\caption{Quantum Amplitude Estimation (QAE) with intermediate states}
\label{alg:qae_intermediate}
\begin{algorithmic}[1]
\Require State-preparation unitary $A$, controlled Grover iterate $\Lambda(Q)$, number of phase qubits $n$ ($N=2^n$)
\Ensure Estimate $\tilde{a}$ of the target amplitude $a$

\State Allocate two $n$-qubit registers: phase register $R$ and system register $S$
\State Initialize the joint state:
\[
\ket{\Gamma_1} \gets \ket{0}_R \otimes A\ket{0}_S
\]

\State Apply the QFT to the phase register:
\[
\ket{\Gamma_2} \gets \mathrm{QFT}_R\,\ket{\Gamma_1}
          \;=\; \frac{1}{N}\sum_{j=0}^{N-1}\ket{j}_R \otimes \ket{\psi}_S
\]

\State Apply $\Lambda(Q)$ to the system:
\[
\ket{\Gamma_3} \gets \Lambda(Q)\,\ket{\Gamma_2}
\]
\[
\ket{\Gamma_3}
= \frac{-i}{\sqrt{2N}}\sum_{j=0}^{N-1}\ket{j}_R\otimes
\Big(e^{(2j+1)i\theta}\ket{\psi_+}_S - e^{-(2j+1)i\theta}\ket{\psi_-}_S\Big)
\]
\[
= \frac{e^{i\theta}}{\sqrt{2}}\ket{S(\theta/\pi)}_R\otimes\ket{\psi_+}_S
  +\frac{e^{-i\theta}}{\sqrt{2}}\ket{S(1-\theta/\pi)}_R\otimes\ket{\psi_-}_S
\]

\State Apply the inverse QFT to the phase register:
\[
\ket{\Gamma_4} \gets \mathrm{QFT}_R^\dagger\,\ket{\Gamma_3}
= \frac{e^{i\theta}}{\sqrt{2}}\ket{N\theta/\pi}_R\otimes\ket{\psi_+}_S
 -\frac{e^{-i\theta}}{\sqrt{2}}\ket{N(1-\theta/\pi)}_R\otimes\ket{\psi_-}_S
\]

\State Measure the phase register $R$ in the computational basis to obtain $y$ (do not measure $S$):
\[
y = N\theta/\pi \quad \text{or} \quad y = N(1-\theta/\pi)
\]

\State Use the sine symmetry to output the amplitude estimate:
\[
\tilde{a} \gets \sin^2\!\left(\pi y/N\right)
\]

\end{algorithmic}
\end{algorithm}

These steps are essentially equivalent to QPE of the $Q$ operator, except the second register is a superposition of eigenstates. Assuming $N\equiv2^n \gg 1$, the error in the estimate for $a$ scales as $|\Tilde{a} -a| \sim k \pi/N$ and exponentially vanishes with increasing number of qubits while using exactly $N$ evaluations of the boolean function $\chi$.

To embed QAE within the PDE solver, recall that the only quantum task is the evaluation of integrals of the form $\int f[\alpha(t)]\,\mathrm{d}t$ over the primary time subintervals. Each such integral is approximated by sampling $f[\alpha(t)]$ at $N$ discrete points within the associated secondary subintervals and forming a Riemann sum i.e., the average of these samples. The problem therefore reduces to estimating the mean of $N$ values,
\begin{equation}\label{e:00}
    \Bar{g}=\frac{1}{N}\sum_{j=0}^{N-1}g(j),
\end{equation}
where $g(j)$ is an affine rescaling of $f[\alpha(t_j)]$ so that $g(j)\in[0,1]$.

To cast mean estimation into a form amenable for QAE, a $(n+1)$-qubit Hilbert space is introduced as the combination of a n-qubit register with computational basis states $\ket{j}$ and a chaperon/ancilla qubit $\ket{z}_c$. The chaperon serves to tag \emph{good} versus \emph{bad} outcomes as \(g(j)\in[0,1]\) is
real valued rather than Boolean.

Assume access to an oracle $\Lambda_2$ that encodes the samples $\{g(j)\}$ as controlled rotations about the ancilla
\begin{equation}\label{e:nv}
    \Lambda_2\ket{j}\otimes\ket{1}_c=\sqrt{g(j)}\ket{j}\otimes\ket{1}_c + \sqrt{1-g(j)}\ket{j}\otimes\ket{0}_c,
\end{equation}
\begin{equation}\label{e:nv1}
    \Lambda_2\ket{j}\otimes\ket{0}_c=\sqrt{1-g(j)}\ket{j}\otimes\ket{1}_c + \sqrt{g(j)}\ket{j}\otimes\ket{0}_c.
\end{equation}
The state preparation unitary $A$ is now defined as 
\begin{equation}
    A = \Lambda_2 \big(F_n \otimes I\big),
\end{equation}
where \(F_n\) is the \(n\)-qubit quantum Fourier transform on the index register and $I$ is the identity operator on the chaperon qubit.
Starting from $\ket{0}^{\otimes n}\ket{1}$,
\begin{equation}
  \ket{\psi}=A\ket{0}^{\otimes n}\otimes\ket{1}_c=\frac{1}{\sqrt{N}}\sum_{j=0}^{N-1}\left(\sqrt{g(j)}\ket{j}\otimes\ket{1}_c + \sqrt{1-g(j)}\ket{j}\otimes\ket{0}_c \right), 
\end{equation}
which after introducing the normalised states 
\begin{equation}
\ket{n_1}=1/\sqrt{N}\sum_{j=0}^{N-1}\sqrt{g(j)/\Bar{g} }\ket{j}\otimes \ket{1}_c \quad, \quad \ket{n_0}=1/\sqrt{N}\sum_{j=0}^{N-1} \sqrt{(1-g(j) )/\Bar{g}}\ket{j}\otimes \ket{1}_c   
\end{equation}
reduces to 
\begin{equation}
    \ket{\psi}=A\ket{0} \otimes \ket{1}_c =\sqrt{\bar g}\ket{n_1}+ \sqrt{1-\bar g}\ket{n_0}.
\end{equation}
This is exactly the canonical QAE form $\ket{\psi}=\sqrt{a}\ket{n_1}+\sqrt{1-a}\ket{n_0}$ with
\(a=\bar g\). Applying amplitude estimation to \(\ket{\psi}\) therefore yields an estimate \(\tilde a\)
of the desired mean \(\bar g\). 
The approach has been applied as a PDE solver to steady state flow through a de Laval nozzle, both with and without shocks \cite{gaitan2020finding}, to nonlinear radiation diffusion of a propagating Marshak wave \cite{gaitan2024simulating}, to tephra dispersal during a submarine volcanic eruption \cite{basu2024quantum} and shows potential for broader classes of nonlinear PDEs \cite{gaitan2021finding}.

From an encoding perspective, the algorithm strategically restructures the Navier-Stokes equations so that the nonlinear evaluation of the driver/integrand is performed classically and only the linear aggregation (time-integration) is delegated to quantum amplitude encoded techniques.
This separation keeps the solver agnostic to the specific PDE and boundary conditions, as different governing equations or discretizations just change the classical values fed to the quantum mean estimator.

Yet, naive amplitude encoding alone suffers from the measurement bottleneck where obtaining precise estimates typically requires a large number of repeated measurements.
The solver circumvents this by incorporating a Grover operator, which moves the desired information into phase differences. These phases are then efficiently extracted using the QPE algorithm. While this approach dramatically reduces the number of measurements needed, the tradeoff is introducing a dependency on register resolution due to the use of basis encoding for phase extraction, limiting accuracy by the number of available qubits for data representation.

A central data representation choice in QCFD is how to encode the spatially discretized fields.
Rather than encoding the entire spatial domain in one large, entangled state as is commonly done in QCFD, this algorithm adopts a localized encoding. Each grid point is treated independently, preparing (and estimating) the local time integral from samples on that point's subintervals.
Such a localized encoding avoids the circuit overheads associated with global entanglement across the mesh, but requires separate quantum queries at each grid point.
Even though this foregoes the exponential compression typically leveraged by amplitude encoding across the spatial domain, it opens up possibilities for parallelism \cite{basu2025solving}. 
As the integration at each grid point is independent, in principle all grid points could be executed simultaneously provided sufficient quantum resources are available. 
Parallelism can also be extended across the subintervals at each grid point, as the function evaluations for different secondary subintervals are mutually independent. Consequently, the quantum states prepared via the oracle (as defined in Equations \ref{e:nv}-\ref{e:nv1}) remain disentangled, ensuring that measurement induced collapse in one subinterval does not affect others.
However, exploiting such parallelism in practice is constrained by the algorithm's hybrid quantum-classical character, where classical functional evaluations, preprocessing and frequent data transfers between classical and quantum systems introduce communication/encoding overheads that could dominate at scale. Careful accounting of these costs is required when comparing wall-clock performance as the original theoretical analysis is in terms of an abstract oracle and the associated query complexity.

Regarding likelihood of quantum speedup, it is important to recall that the underlying driver function $f(U)$ in Equation~\ref{e:6}) is assumed to belong to a H\"older class, characterized by a certain order of differentiability and a H\"older continuity exponent, which together quantify the function's smoothness.
Kacewicz's complexity analysis suggested that a quadratic quantum speedup in query complexity can be achieved for such functions, but notably only when the driver is \emph{not} smooth. 
Although such non-smooth cases may still arise in practical applications, the broader implications of this limitation for more general or smooth drivers remain an open question and warrant further study. 

As mentioned previously, the quantum speedup is expressed in terms of oracle query complexity. While oracle-based models are standard in quantum algorithm analysis, they abstract away the actual cost of implementing these oracles as quantum circuits. In practice, the quantum cost of evaluating an oracle can differ substantially from its classical counterpart, especially when complex or deep reversible arithmetic is involved.
Therefore, to claim a \emph{practical} quantum advantage, it is essential to go beyond the oracle model and analyze the circuit-level complexity of encoding and evaluating each oracle. 
A first step in this direction was made in a subsequent work \cite{gaitan2024circuit} by constructing explicit circuits for each of the quantum oracles used in the algorithm (i.e. $Q,\Lambda_1, \Lambda_2$). 
However, even though these implementations represent progress, further refinement is needed to reduce circuit depth and gate count, leaving open the question of whether the theoretical speedup translates into tangible computational gains.

Additionally, the solver relies on QAE, which carries a non-zero probability of failure i.e. returning an estimate outside the desired error tolerance. 
Prior work \cite{heinrich2002quantum} has demonstrated that by running QAE $M$ times and returning the median value of the results, it is possible to control the failure probability as $\delta=\exp[-M/8]$. So although it is possible to retain some control over the success rate this comes at the cost of repeated samples, each with its own computational cost that should be accounted for. 
Similarly, whether precision limitations have an impact on the fact that the driver function has to be rescaled ($f \rightarrow g$) for QAE and then scaled back for classical processing is not entirely clear, particularly for cases where the function spans a wide dynamic range. 

\subsection{Simulating Nonlinearity With Interacting Copies}\label{sec:2}
The previous section described a common strategy for handling nonlinearity in QCFD under the limitations of quantum encoding, by reformulating the system so that only linear operations are delegated to the quantum computer. 
An alternative appraoch is to simulate nonlinearity directly by exploiting interactions between multiple quantum registers. While true nonlinear operations on amplitudes are unlikely to be possible in quantum mechanics, coordinated operations across multiple copies can effectively emulate nonlinear behavior.

The first such construction, detailed in \cite{leyton2008quantum}, considers $n$ first-order nonlinear ODEs defined by polynomials $f_\alpha(\mathbf{z})$ in $n$ variables $z_j$ ($j=1,\dots,n$)
\begin{equation}\label{e:7}
\frac{dz_j}{dt}=f_j\big(z_1(t),\dots,z_n(t)\big),\qquad \mathbf{z}(0)=\mathbf{b}.
\end{equation}
Variables are amplitude-encoded into an $(n{+}1)$-level system as
\begin{equation}
\ket{\phi}=\frac{1}{\sqrt{2}}\ket{0}+\frac{1}{\sqrt{2}}\sum_{j=1}^n z_j\ket{j},\qquad \sum_{j=1}^n|z_j|^2=1,
\end{equation}
noting that any hardware realization still uses a conventional 2-level (i.e. qubit) encoding.

Consider first the case of a quadratic transformation in amplitudes, with higher orders extending similarly. Such a transformation cannot be enacted with only a single copy but is essential for time-stepping a nonlinear system. The core insight is that given two copies, the combined tensor product becomes
\begin{equation}
\ket{\phi}\ket{\phi}=\frac{1}{2}\sum_{j,k=0}^n z_j z_k \ket{jk},\qquad z_0=1,
\end{equation}
which contains all monomials of degree $\le 2$.  
The goal is isolating and manipulating the desired combinations from this tensor state, which is formalized as the transformation
\begin{equation}\label{e:10}
    \mathbf{z} \rightarrow \mathbf{F}(\mathbf{z}),
\end{equation}
with
\begin{equation}
\mathbf{F}(\mathbf{z})=
\begin{bmatrix}
f_1(\mathbf{z}) \\
f_2(\mathbf{z}) \\
\vdots \\
f_n(\mathbf{z})
\end{bmatrix},
\end{equation}
where each $f_\alpha$ is a quadratic polynomial such that 
\begin{equation}
    f_\alpha(\mathbf{z})=\sum_{k,l=0}^na_{kl}^{(\alpha)}z_k z_l.
\end{equation}
Implementing this transformation is equivalent to creating the state
\begin{equation}\label{e:11}
    \ket{\phi'}=\frac{1}{\sqrt{2}}\sum_{\alpha=0}^nf_\alpha(\mathbf{z})\ket{\alpha},
\end{equation}
where $a_{kl}^{(\alpha)}=a_{lk}^{(\alpha)}$ and $f_0(\mathbf{z})=1$. 
Iterating such a map allows the evolution of the original system of ODEs. 
For example, the forward Euler method reduces to $\ket{\phi(t+\delta t)}=\ket{\phi(t)}+\delta t\ket{\phi'(t)} +O(\delta t^2)$. While this approach faces several important caveats regarding efficiency and scalability, we first describe the methodology.

In order to construct the state in Equation \ref{e:11}, we can define the operator 
\begin{equation}\label{e:8}
    A=\sum_{\alpha,k,l=0}^n a_{kl}^{(\alpha)}\ketbra{\alpha0}{kl },
\end{equation}
which mixes the relevant monomials from the tensor product of copies. However, as this cannot be applied directly, a pointer qubit ($P$) is adjoined with the operator to form the Hamiltonian
\begin{equation}
    H=-iA \otimes \ket{1}_P\bra{0} +i A^\dagger \otimes\ket{0}_P\bra{1}.
\end{equation}
Now preparing the initial state 
\begin{equation}
    \ket{\phi}\ket{\phi}\ket{0},
\end{equation}
and evolving under the Hamiltonian for time $t=\epsilon$ gives
\begin{equation}\label{e:9}
    \ket{\Psi}=e^{i\epsilon H}\ket{\phi}\ket{\phi}\ket{0}=\sum_{j=0}^\infty\frac{(i \epsilon H)^j}{j!}\ket{\phi}\ket{\phi}\ket{0}=\ket{\phi}\ket{\phi}\ket{0} + \epsilon A \ket{\phi}\ket{\phi}\ket{1}\dots ,
\end{equation}
where by construction
\begin{equation}
    A \ket{\phi}\ket{\phi}=\frac{1}{2}\sum_{\alpha , k ,l=0}^n a_{kl}^{(\alpha)} z_kz_l \ket{\alpha}\ket{0}= \frac{1}{\sqrt{2}}\ket{\phi'}\ket{0}.
\end{equation}

Thus, post-selection on the pointer qubit in Equation \ref{e:9} has a probability of approximately $\frac{1}{2}\epsilon^2$ of returning the desired state. This assumes the transformation is measure preserving ($\sum_{j=1}^n|z_j|^2=\sum_{\alpha=1}^n f_\alpha^*(\mathbf{z})f_\alpha^*(\mathbf{z})=1$). Otherwise, although the algorithm remains unchanged, there is a proportional reduction in success probability. 
In the event of failure, it may be possible to recover an earlier intermediate state from the partially measured, poisoned state, but it is not always clear how this can be done in practice. 
Consequently, the original work recommended using $16/\epsilon^2$ fresh pairs of copies to ensure at least one success with high probability. 
A key limitation is that each time step depends on the successful preparation of the previous step's state. 
Moreover, due to the no-cloning theorem it is not possible to generate copies on the fly and so all the required copies must be maintained throughout the algorithm.  
Thus, iterating the map in Equation \ref{e:10} for $m$ steps would require $(16/\epsilon^2)^m$ initial pairs.
The algorithm also scales exponentially with both the inverse Euler step size and the degree of nonlinearity, making it somewhat feasible only for well-conditioned systems.

Although the original study suggests an exponential advantage in resource scaling with the number of variables, this claim rests on the assumption that $\ket{\phi}$ can be prepared efficiently, which as discussed earlier, may not hold in QCFD. Moreover, even if state preparation and controlled evolution without excessive error accumulation were practical, physically meaningful CFD observables must still be efficiently extracted from amplitude encoded quantum states.  
Nevertheless, the approach represents an important conceptual breakthrough by exploiting multiple quantum state copies and nonunitary, destructive measurements to induce effective amplitude nonlinearities, attempting to overcome a fundamental obstacle in leveraging amplitude encoding for QCFD. 

A subsequent work \cite{lloyd2020quantum} reduced the exponential growth of resources with integration time to quadratic, while retaining logarithmic scaling in the state space dimension via amplitude encoding. The approach again leverages multiple copies of the system, but now combined with methods for simulating nonlinear Schr{\"o}dinger-type dynamics and quantum linear system solvers. 
The nonlinear Schr\"odinger equation is a mean-field description in which the apparent nonlinearity of a single subsystem is induced by weak interactions with many identical copies, so that the global evolution still remains linear. 

The dynamics of interest are more conveniently cast here as
\begin{equation}\label{e:12}
    \frac{d x}{dt}+ f(x)x=b(t),
\end{equation}
where $f(x) \in \mathbb{C}^{d \times d}$ is an order $m$ polynomial in the components of $x\in C^d$. Increasing the dimensionality of $x$ by one, allows expressing the nonlinearity as 
\begin{equation}
    f(x)=x^{\dagger\otimes m}Fx^{\otimes m},
\end{equation}
for a tensor $F$ that is assumed to be sparse and with readily computable entries. This could be the case where the underlying physics is dominated by local interactions.

The special case where $f(x)$ is anti-Hermitian and in the absence of driving terms ($b(t)=0$) is particularly suited for quantum simulation, as it allows directly embedding the function into the Hamiltonian
\begin{equation}
    H=-i \binom{n}{m}^{-1}\sum_{j_1 \dots j_m}F_{j_1 \dots j_m},
\end{equation}
where $F_{j_1 \dots j_m}$ denotes $F$ acting on the $m$ distinct copies indexed by ($j_1 \dots j_m$). 
This Hamiltonian is then applied to an initial system consisting of $n$ copies of the state i.e. $x(0)^{\otimes n}$. 
The short time behavior of this larger system is described by the n-system linear Schr\"odinger equation, 
\begin{equation}\label{e:14}
    e^{-iH\Delta t}x^{\otimes n}=(I - iH\Delta t - (1/2)H^2\Delta t^2 + O(\Delta t^3))x^{\otimes n } .
\end{equation}
Moreover, the short time behavior of any single copy is obtained by tracing out the other copies to give
\begin{equation}\label{e:13}
    x \rightarrow (I- \Delta t f(x) )x + O(E^2\Delta t^2).
\end{equation}
Note that throughout, $|E|^2$ represents a time-averaged measure of the squared instantaneous rates set by $f(x)$ and defined by its eigenvalues. In the anti-Hermitian case, this reduces to the square of the time-averaged operator norm. 
Thus, each individual copy undergoes a short time evolution under the desired non-linear Equation \ref{e:12} to first order. The number of Trotter steps, $T$, to evolve the nonlinear Schr\"odinger equation must be chosen so as to keep $E^2\Delta t^2 T=E^2t\Delta t=E^2 t^2/T$ small and as such scales quadratically with respect to integration time ($t$).   

Comparing the full multi-copy Trotterised evolution of Equation \ref{e:14} with the single-copy nonlinear Euler product term by term, reveals how deviations from the latter arise due to higher order terms in the former. 
Specifically, second order terms in the Hamiltonian evolution will include sums over pairs of interaction terms (e.g. $F_SF_S'$). If these tensors are acting on disjoint sets of subsystems ($S \cap S' = \emptyset$) then the other systems effectively factor out when partial tracing to give the desired single state non-linear evolution. 
However, when there are overlapping copies ($S \cap S' \neq \emptyset$), the operators do not commute and create unwanted entanglement across the copies. This leads to partial tracing resulting in a mixed state rather than the desired pure state.
The fraction of terms with this unwanted overlap are suppressed by a factor of ($1/n$) compared to the numerical discretisation error, provided that $n \gg mT$ and so the required number of copies, similarly to the Trotter steps, also scales quadratically in time.   

While realistic for applications such as the Gross-Pitaevski equation, in the context of QCFD the requirement of $f(x)$ being anti-Hermitian is not guaranteed. Simulating the Navier-Stokes and other fluid dynamic descriptions requires extending the methodology to account for more general nonlinear differential equations as $f(x)$ is no longer anti-Hermitian and cannot be directly embedded into a Hamiltonian.

As detailed in \cite{lloyd2020quantum}, a potential solution is reconfiguring the problem into a form amenable for quantum linear system solvers. Note that although the state remains normalized under the unitary evolution of anti-hermitian $f(x)$, maintaining normalization throughout the integration period under the non-Hermitian operator of Equation \ref{e:15} requires rescaling variables and adding an extra dimension which is not described here. 
Taking $n$ copies of the input state and adjoining the system with a time-step register ($\ket{k}$) gives
\begin{equation}
 \ket{B}^{(n)} = \ket{b_0}^{\otimes n}\ket{k=0} + \Delta t \sum_{k=1}^T\ket{b_k}^{\otimes n}\ket{k},
\end{equation}
where $k$ indexes the corresponding time step. 
The dynamics of a nonlinear Schr\"odinger equation can then be encapsulated into the operator
\begin{equation}\label{e:15}
    \mathcal{M}^{(n)}= \sum_{k=0}^{T}I \otimes \ket{k}\bra{k}-\sum_{k=0}^{T-1}(I-\Delta t \binom{n}{n}^{}\sum_{j_1 \dots j_m}F_{j_1 \dots j_m})\otimes \ket{k+1}\bra{k}.
\end{equation}
Together this forms a linear system problem of the form
\begin{equation}
    \mathcal{M}^{(n)}\ket{X}^{(n)}=\ket{B}^{(n)},
\end{equation}
where, subject to the error in the nonlinear Schr\"odinger equation approximation, the solution is the history state
\begin{equation}
    \ket{X}^{(n)} \approx \sum_k \ket{x_k^1}\otimes \dots \otimes \ket{x_k^n}\ket{k}.
\end{equation}
With each copy representing the forward Euler solution of Equation \ref{e:12} at time $k$,
\begin{equation}
    \ket{x_k}=(I-\Delta t f(x_{k-1}))\ket{x_{k-1}} + \Delta t \ket{b_k}.
\end{equation}
As in the anti-Hermitian case, the dominant contribution to the total error arises from discretisation in the numerical integration, scaling as 
$|E|^2 t\,\Delta t$, while deviations from the nonlinear \"odinger mean-field approximation are smaller by a factor of $1/n$, scaling as 
$|E|^2 t\,\Delta t\, m^2 / n$. In the present setting of a non-Hermitian Hamiltonian, $|E|^2$ is defined as the time-averaged modulus squared of the complex eigenvalues of $f(x)$ over the integration interval. 

Discretisation errors could in principle be reduced by using higher-order Trotterization schemes or more sophisticated numerical integration methods. However, the core approximations underlying the nonlinear quantum solver inevitably breaks down for sufficiently strong nonlinearities. In particular, if the nonlinear system exhibits positive Lyapunov exponents and the integration time is long enough that the stability constraint ($|E|^2t \Delta t < O(\epsilon) $) is violated, the accumulated forward-Euler discretisation error will grow beyond the target accuracy. Crossing this threshold would, in effect, allow the amplification of exponentially small differences in the initial state, implying that quantum computers could efficiently solve NP-complete problems, which is widely believed to be impossible.

This limitation is especially relevant for nonlinear PDEs such as those in fluid dynamics, where strong instabilities, turbulence, or chaotic behavior can drive rapid growth in $|E|$. Understanding precisely where the method remains stable and efficient therefore requires a careful analysis of the interplay between nonlinearity strength, the Lyapunov spectrum and numerical discretisation, particularly for challenging regimes like high Reynolds number turbulence.
This is also true for the approach proposed in \cite{liu2021efficient} which similarly created a linear system over multiple copies, but using Carleman linearisation as opposed to Schr{\"o}dinger linearisation. Where the algorithm was shown to also scale almost quadratically in evolution time but only for certain flow cases, categorized by weak nonlinearity and forcing compared with linear dissipation. 

Note that some formulations of fluid dynamics might be more amenable than others to linearisation and subsequent application of a quantum linear system solver. One example is the Lattice Boltzmann Equation \cite{li2025potential} where the nonlinearity is formally governed by the Mach number as opposed to the Reynolds number - a significantly less restrictive condition. 
Notwithstanding this, all these approaches fundamentally inherent both the strengths and weaknesses of the quantum linear system solver. While this may hold the promise of exponential speedup in theory, leveraging this advantage in practice is subject to very strong caveats \cite{aaronson2015read}. Most notable perhaps is the challenge of obtaining meaningful measurements from an amplitude encoded quantum state efficiently, which as discussed earlier is far from trivial in the context of QCFD. 
\subsection{Optimization Using Quantum Annealing}\label{sec:3}
An alternative paradigm for quantum fluid simulation is to reformulate the problem as an optimization task and solve it using quantum annealing (QA) \cite{finnila1994quantum,kadowaki1998quantum}. QA is closely related to adiabatic quantum computing (AQC) \cite{albash2018adiabatic,aharonov2008adiabatic}, but is better understood as a heuristic quantum optimization framework rather than a universal model of computation. AQC relies on the adiabatic theorem, which guarantees that a quantum system initialized in the ground state of a Hamiltonian will remain in its instantaneous ground state during a slow, continuous evolution, provided that the Hamiltonian changes slowly enough and that a finite spectral gap separates the ground and excited states. 
By contrast, QA relaxes the strict requirement of adiabaticity, by no longer requiring a prohibitively slow Hamiltonian schedule, although consequently the system may undergo non-adiabatic transitions. As a result, QA trades the theoretical guarantees and deterministic convergence of ideal AQC for a probabilistic, heuristic sampling procedure in which unwanted excited states may also be sampled. Moreover, QA behaves as a quantum-enhanced optimizer that can exploit phenomena such as quantum tunneling to escape local minima, potentially outperforming classical heuristics. 

Quantum annealing has been explored for a variety of combinatorial optimization problems \cite{yarkoni2022quantum} including scheduling, quantum chemistry, machine learning and material design, with mixed but promising results. A comprehensive discussion of QA and its theoretical foundations can be found in the extensive pedagogical literature \cite{rajak2023quantum,hauke2020perspectives}. For the purposes of this work, it is sufficient to appreciate that QA encodes candidate solutions as computational basis states representing bit-strings that minimize an objective function. 
As such, it does not directly exploit the exponential representational capacity of amplitude encoding. Nevertheless, it can offer performance benefits for certain classes of problems due to its ability to traverse complex energy landscapes efficiently \cite{abel2021quantum,abel2022quantum,farhi2002quantum}.  

Currently, quantum annealing is primarily viable for combinatorial problems that can be formulated in a manner compatible with existing annealing hardware, such as quadratic unconstrained binary optimization (QUBO) problems or their equivalent Ising Hamiltonian representations \cite{lucas2014ising}. 
This is also the case for D-Wave quantum annealers, which remain the most mature commercial QA systems available today. 
Consequently, a significant secondary challenge arises in reformulating fluid dynamics problems into QUBO or Ising form, a process that often requires substantial restructuring and approximation of the original problem.

Several efforts have been made in this direction, including formulations for simple channel flow \cite{ray2022viability}, the development of auxiliary routines to support fluid solvers \cite{rathore2025load}, applications to aerospace design optimization \cite{kuya2025quantum}, the advection diffusion equation \cite{takagi2024implementation} and a lattice gas implementation \cite{kuya2024quantum}. 
In this work, we restrict our attention to two representative approaches for encoding fluid dynamics into QUBO/Ising form. 
Namely, a full problem encoding, where the entire fluid dynamics problem is mapped onto the quantum annealer and a hybrid encoding, where only a computationally intensive subroutine or bottleneck is offloaded to the annealer.

\subsubsection{Quantum Annealing For Partial Differential Equations}
Recently, a general-purpose method for using quantum annealer to solve coupled partial differential equations named Qade was proposed \cite{criado2022qade}. Its main advantage over prior endeavours \cite{srivastava2019box,zanger2021quantum} is inherent flexibility resulting from making no assumptions regarding the underlying equations beyond linearity in the unknown fields and their derivatives. 
Motivated by machine learning primitives, the PDEs are written in functional form
\begin{equation}
    E_i(x)[f]|_{x \in \mathcal{X_i}} = 0 ,
\end{equation}
where $E_i(x)$ are local functionals of the field $f$, enforced over domains $\mathcal{X}_i$. Boundary and initial conditions are straightforwardly incorporated as additional functionals within the set. 
Assuming linear dependence on $f$ and its derivatives 
\begin{equation}
    E_i(x)[f]=\sum C_{in}^{(k)}(x) \cdot (\partial^kf_n(x))+B_i(x),
\end{equation}
with inhomogeneous terms $B_i(x)$, variable coefficients $C_{in}^{(k)}$ and derivatives up to order $k$.  

Discretising each domain onto a finite subsets of sample points ($X_i \subset \mathcal{X_i}$) yields the loss function
\begin{equation}
    L[f]=\sum_i\sum_{x\in X_i}(E_i(x)[f])^2,
\end{equation}
which has a global minimum corresponding to satisfying all equations across each of the sample points. 
To keep resources tractable, $f$ is parameterized as a finite expansion in basis functions, $\Phi_m$, as 
\begin{equation}\label{e:21}
    f_n(x)=\sum_mw_{nm}\Phi_m(x).
\end{equation}
The equations can now be expressed in terms of this finite set of weight parameters, $w_{nm}$, as 
\begin{equation}
    E_i(x,w)=\sum_{nm}H_{in}(x)[\Phi_m]w_{nm} + B_i(x),
\end{equation}
 \begin{equation}
     H_{in}(x)[\Phi]=\sum_k C_{in}^{(k)}(x) \cdot (\partial^k\Phi(x)),
 \end{equation}
which give the quadratic loss function 
\begin{equation}\label{e:20}
    L(w)=\sum_{nm \, pq}w_{nm}J_{nm,pq}w_{pq} + \sum_{nm}h_{nm}w_{nm},
\end{equation}
with
\begin{equation}
    J_{nm, pq}=\sum_i \sum_{x\in X_i}H_{in}(x)[\Phi_m]H_{ip}(x)[\Phi_q],
\end{equation}
\begin{equation}
    h_{nm}=2 \sum_i \sum_{x\in X_i}H_{in}(x)[\Phi_m] B_i(x) . 
\end{equation}

To map this to a binary Ising/QUBO, each continuous weight is encoded in spins, $\hat{w}^{(\alpha)}_{nm}=\pm1$, as 
\begin{equation}\label{e:22}
    w_{nm}=c_{nm}+s_{nm}\sum_{\alpha=1}^{n_{spins}}\frac{\hat{w}_{nm}^{(\alpha)}}{2^\alpha}
\end{equation}
where the free parameters $c_{nm}$, $s_{nm}$ center and scale the dynamic range. 
Substituting this encoding into the continuous quadratic loss function of Equation \ref{e:20} yields the final Ising objective 
\begin{equation}
    L(w)=\sum_{nm \, pq \alpha \beta}\hat{w}^{(\alpha)}_{nm}\hat{J}^{(\alpha\beta)}_{nm,pq}\hat{w}^{(\beta)}_{pq} + \sum_{nm\alpha}\hat{h}^{(\alpha)}_{nm}\hat{w}^{(\alpha)}_{nm},
\end{equation}
with
\begin{equation}
    \hat{J}^{(\alpha \beta)}_{nm, pq}=2^{-(\alpha +\beta)}s_{nm}s_{pq}J_{nm,pq},
\end{equation}
\begin{equation}
    \hat{h}^{(\alpha)}_{nm}=2^{-\alpha}s_{nm}(h_{nm} + 2c_{pq} J_{nm,pq}) . 
\end{equation}
This can now be solved on a quantum annealer, with the resulting bitstrings decoded into weights using Equation \ref{e:22} and the approximate solution reconstructed using Equation \ref{e:21}. 

A key practical limitation is the number of available spins/qubits. The total number of spins scales as the product of (i) spins per weight, (ii) the number of basis functions, and (iii) the number of unknown fields being solved for. On current hardware this quickly becomes prohibitive as any one of these factors grows. 
To mitigate this, iterative strategies effectively \emph{zoom-in} on the solution \cite{criado2022qade,zlokapa2020quantum} by updating the encoding parameters $c_{nm}$ and $s_{nm}$ in Equation~\ref{e:22} after each iteration, re-centering the encoding around the current estimate and reducing the scale to improve effective precision with fewer spins.
This helps reduce the spin footprint, though it is important to monitor for error accumulation as early biases can persist and be amplified across epochs.

\begin{figure}[t]
    \centering
    \includegraphics[width=0.95\linewidth]{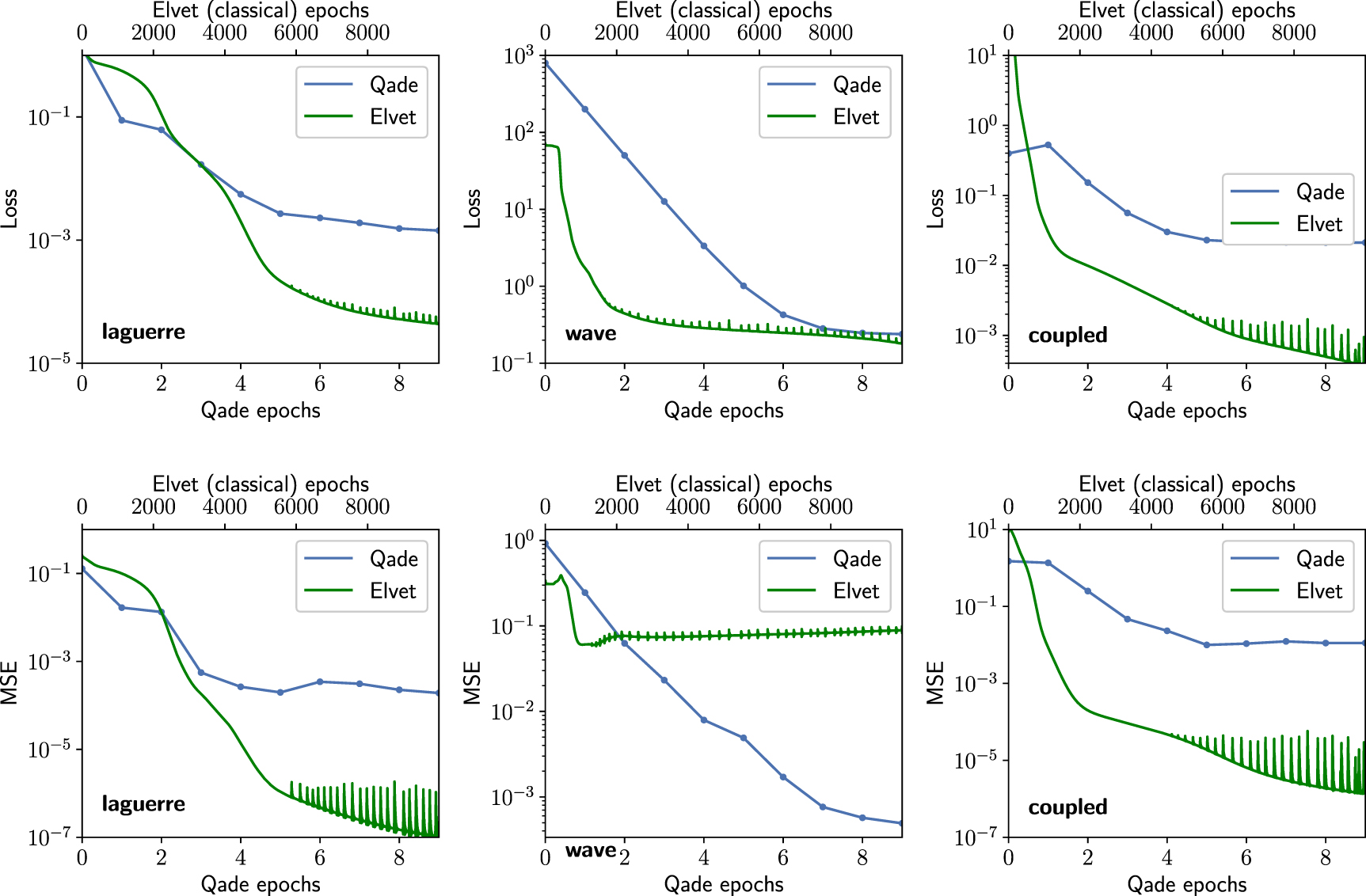}
    \caption{Performance comparison between Qade and the state of the art, neural network based classical solver, Elvet, for several differential equations. The figure reports the loss and mean squared error (MSE) as functions of training epoch. The results are problem-dependent: while Elvet ultimately achieves higher accuracy in two of the three cases, Qade performs especially well for the wave equation example, where it attains a lower MSE and a comparable final loss. Adapted from \cite{criado2022qade} (\href{https://creativecommons.org/licenses/by/4.0/}{CC BY}).}
    \label{fig:3}
\end{figure}

The Qade framework makes it comparatively straightforward to encode a broad class of linear PDEs for annealing. Demonstrations include the Laguerre equation, the wave equation, coupled PDEs \cite{criado2022qade} as illustrated in Figure \ref{fig:3}, and oceanographic systems \cite{matsuta2025formulation}, as well as initial explorations of the Navier-Stokes albeit restricted to only the continuity equation \cite{rodriguez2024quantum}.
Because QA lacks the adiabatic guarantees of adiabatic quantum computing, provable and guaranteed speedups are elusive. Even so, empirical results on hard optimizations suggest QA can sometimes deliver higher quality solutions: an expectation likely to strengthen as hardware matures. 
More broadly, constrained continuous variable optimization is notoriously difficult \cite{murty1985some}, even with only seemingly simple constraints such as positivity, so there is some reason to expect QA might be well placed to avoid poor local minima that would otherwise hinder classical heuristics. In practice, solution quality is not guaranteed and is highly sensitive to annealing hyperparameters including schedule design, number of reads, problem embedding, and chain strengths. These parameters are difficult to tune \emph{a priori} and so careful calibration, problem specific heuristics, as well as validation against classical baselines remain essential.

From an encoding standpoint, the initial state is easy to prepare (an equal superposition). Bitstring readout also simplifies measurement relative to amplitude encoded, gate model approaches, but annealing remains a heuristic sampler where low energy samples are not guaranteed to include the true optimum, and post-selection still matters. Equation \ref{e:22} highlights an intrinsic challenge in representing real valued coefficients with few binary spins. The choice of $c_{nm}$, $s_{nm}$ and bit depth directly controls dynamic range and precision. Recently proposed novel real number encodings for annealers \cite{endo2024novel} may offer better tradeoffs and integrating such schemes into this framework is a natural direction for future work. 
Moreover, even though basis function expansions keep the dimensionality manageable, efficiency relies on good \emph{a priori} basis selection. Without domain knowledge (or adaptive basis refinement), the method can remain valid but might not remain efficient.

Lastly, extending the framework to capture nonlinearity is essential for broader CFD applications. In the annealing setting, the issue is not that all nonlinear structure is excluded outright, since the native quadratic couplings are already nonlinear in the encoded variables. Rather, the challenge is to determine which forms of nonlinearity can be represented naturally and efficiently within the annealing Hamiltonian. In this context, unary type encodings may be particularly useful, since they allow arbitrary local and pairwise interactions to be expressed using only quadratic Ising terms. Whether these classes of interaction may already capture many of the most important nonlinear structures relevant to fluid modelling is an open question.

More general nonlinear terms may still require the introduction of auxiliary variables together with penalty constraints, followed by quadratization into a QUBO form. This preserves compatibility with the annealing workflow, but typically increases the number of spins and the connectivity required, so the resulting trade-offs must be assessed carefully. Related ideas have already been explored in other areas, for example in the encoding of strongly nonlinear local terms arising in quantum field theory on annealing-type hardware \cite{abel2021quantum}. It may therefore be interesting to investigate whether similar constructions can be adapted to fluid problems. As hardware capabilities continue to improve, quantum annealing may provide a practical route to certain high-dimensional nonlinear problems, particularly where the dominant interactions can be represented in a form well matched to the native structure of the device.

\subsubsection{Quantum Runge-Kutta Methods}
Many quantum algorithms adopt simple time integrators to evolve differential equations, such as the forward Euler method used in Section \ref{sec:2}. In realistic CFD, however, higher order schemes with better stability and accuracy are typically preferred. The Runge-Kutta (RK) framework unifies a broad family of explicit and implicit methods under a single formulation.

Consider the initial value problem
\begin{equation}\label{e:23}
  \frac{d}{dt}u(t)=f(u(t),t),\qquad u(t)\in\mathcal{R}^N,\quad u(0)=u_0.
\end{equation}
An $s$-stage RK method is specified by a Butcher tableau $(A,b,c)$ with $A\in\mathbb{R}^{s\times s}$ and $b,c\in\mathbb{R}^s$ \cite{butcher1963coefficients}. A single time update is given by 
\begin{equation}
    \Tilde{u}(t+\Delta t)= \Tilde{u} + \Delta t \sum_{i=1}^s b_ik_i,
\end{equation}
where the stage values satisfy
\begin{equation}
    k_i=f(\Tilde{u}(t) + \Delta t \sum_{j=1}^s A_{ij}k_j,t+\Delta tc_i, \qquad i=1,\dots,s.
\end{equation}
The method is explicit when $A$ is strictly lower triangular, so each $k_i$ depends only on $k_1,\dots,k_{i-1}$. Otherwise, the method is implicit and the stages are coupled.

For the large, stiff systems commonly found in CFD, resolving many implicit stages is computationally demanding. Classical implementations typically employ iterative methods to converge the coupled stage equations to within an acceptable tolerance \cite{dutt2000spectral,emmett2012toward}. 
These iterations can quickly become costly, particularly for higher order methods and so this stage-solve can be a viable target for acceleration within algorithms that offload this bottleneck to quantum computers.

A method to reframe RK integration into a form amenable to quantum annealears was detailed in \cite{zanger2021quantum} and is summarised here. Assuming a polynomial system, Equation \ref{e:23} can be written as 
\begin{equation}
  \frac{d}{dt}u(t)=f(u(t),t)= \sum_{i=1}^M L^{(i)}u(t)^{\otimes(i-1)},
\end{equation}
with $u(t)\in\mathcal{R}^N,u(0)=u_0$ and $L^{(i)} \in (\mathcal{R^N)^{\otimes i}}$ where $M$ is the highest degree polynomial. 

The RK time step is thus
\begin{equation}\label{e:24}
    \Tilde{u}_j (t+ \Delta t) = \Tilde{u}_j + \Delta t \sum_{o=1}^sb_0K_{oj},
\end{equation}
for stages
\begin{equation}\label{e:25}
K_{o j}=
\underbrace{
\sum_{i=1}^{M}
L^{(i)}_{j k_1\dots k_{i}}
\tilde{u}_{k_1}(t)\dots \tilde{u}_{k_{i}}(t)
}_{\text{evaluate } f(\tilde{u}(t)) \text{ in components}}
+ \Delta t
\underbrace{
\sum_{e=1}^{s} A_{o e}
\sum_{i=1}^{M}
L^{(i)}_{j k_1\dots k_{i}}
K_{e k_1} \dots K_{e k_{i}}
}_{\text{implicit RK coupling via other stages}}
,
\end{equation}
with $k_1 \dots k_i \in \{1, \dots N\}$ and $K \in \mathcal{R}^{s \times N}$ .

This step is recast as an unconstrained least-squares problem by taking the sum of squared residuals of Equations~\ref{e:24} and~\ref{e:25} simultaneously, so that both the coupled stage equations and the final update are enforced in one objective. Optimizing over the stage values and the next step state yields
\begin{equation}
\begin{aligned}
\tilde{u}_j(t_{m+1})
&= \arg\min_{\tilde{u}_j(t_{m+1})}\;
\min_{K_1,\ldots,K_s}
\Bigg[
\left(
\tilde{u}_j(t_{m+1}) - \tilde{u}_j(t_m) - \Delta t \sum_{o=1}^{s} b_o K_{o j} \right)^2
\\
&
+\sum_{o=1}^{s} \left(
K_{o j}- \sum_{i=1}^{M} L^{(i)}_{j k_1\dots k_{i}}\, \tilde{u}_{k_1}(t_m)\dots \tilde{u}_{k_{i}}(t_m)
-
\Delta t \sum_{e=1}^{s} A_{o e}
\sum_{i=1}^{M}
L^{(i)}_{j k_1\cdots k_{i}}\,
K_{e k_1}\cdots K_{e k_{i}}
\right)^2
\Bigg].
\end{aligned}
\end{equation}

The optimization variables are $\tilde{u}_j(t_{m+1})$ and the stage entries $K_{o j}$, all of which are continuous and real. To convert this continuous problem into a QUBO, these real numbers must be encoded in a binary representation. However, the required connectivity of underlying hardware scales drastically as the number of bits per real number approximation increases, since each extra bit introduces additional couplers that must be embedded on sparse device graphs. 

To mitigate this, \cite{zanger2021quantum} proposes an adaptive, windowed encoding. A real number, $g$, is represented as
\begin{equation}
    g \approx 2^{n-1-k} \sigma_{n-1}+\dots 2^{0-k} \sigma_0 +d = 2^{-k}\sigma + d, 
\end{equation}
where, during an anneal, only the binary vector $\sigma$ is optimized. The scale $k$ and offset $d$ are treated as classical parameters and updated between anneals to refine the window for the next iteration. This yields a hybrid variational loop in which the quantum subroutine is complemented by iterative classical post-processing.
The scheme reduces the qubit footprint and, for convex objectives (typically arising from linear dynamics), effectively decouples numerical precision from the number of bits used per variable. In such linear settings the method performs well and exhibits favourable scaling with system size. The anneal optimizes all variables simultaneously, so the cost of increasing problem dimension is expressed primarily in concurrent hardware resources rather than additional sequential time, as would be the case for many classical solvers.

Nonlinearity, however, reintroduces significant challenges once again. Enforcing polynomial interactions requires reduction by substitution \cite{tanahashi2019application}, which introduces auxiliary spins that can grow exponentially with the polynomial order $M$. Moreover, the outer variational loop is no longer guaranteed to converge in the nonconvex/nonlinear case \cite{zanger2021quantum}, with reported discrepancies for a simple Riccati equation illustrating this limitation. While future hardware with lower noise and more qubits may alleviate some issues, it is also natural to explore alternative encodings that minimize connectivity requirements. For example, domain wall encodings \cite{chancellor2019domain} can replace demanding long range couplers with short range chains at the expense of additional qubits, potentially improving embedding quality on near term architectures without excessive reliance on the variational strategy. 

\subsection{Quantum lattice Boltzmann Method}\label{sec:4}
The Quantum Lattice Boltzmann Method (QLBM) is a quantum adaptation of the classical Lattice Boltzmann Method (LBM) \cite{kruger2017lattice}, developed to simulate fluid dynamics and related physical systems on quantum computers. 
In a classical setting, the LBM is widely used in CFD as an alternative to directly solving macroscopic partial differential equations. Instead, it models the evolution of mesoscopic distribution functions, which encode the probability of finding particles at a given location with a specified velocity. Macroscopic quantities of interest (such as density or momentum) emerge naturally as moments of these distributions. Note that readers already familiar with LBM for the advection-diffusion equation may choose to skip ahead to Equation \ref{e:lbm}.

At the kinetic level, the governing description is provided by the Boltzmann equation,
\begin{equation}
    \frac{\partial f}{\partial t} + \mathbf{v} \cdot \nabla_{\mathbf{x}} f = \Omega(f),
\end{equation}
where $f(\mathbf{x}, \mathbf{v}, t)$ is the single-particle distribution function, $\mathbf{v}$ denotes the \emph{microscopic particle velocity} variable in phase space (distinct from the macroscopic fluid velocity) and $\Omega(f)$ is the collision operator describing local particle interactions. 
Discretizing space, time, and velocity space leads to the lattice Boltzmann formulation,
\begin{equation}\label{e2222}
    f_i(\mathbf{x} + \mathbf{c}_i \Delta t, t + \Delta t) 
    = f_i(\mathbf{x}, t) + \Omega_i\left(f(\mathbf{x}, t)\right).
\end{equation}
where $f_i$ is the discrete distribution associated with velocity $c_i$.
A common simplification is the Bhatnagar-Gross-Krook (BGK) model \cite{bhatnagar1954model}, where the collision operator relaxes the system toward a local equilibrium $f_i^{eq}$
\begin{equation}\label{e:16}
    f_i(\mathbf{x} + \mathbf{c}_i \Delta t, \, t + \Delta t) 
    = f_i(\mathbf{x}, t) - \frac{\Delta t}{\tau}\left(f_i(\mathbf{x}, t) - f_i^{eq}(\mathbf{x}, t)\right),
\end{equation}
with $\tau$, the relaxation time, linked to viscosity at the macroscopic scale. Different choices for the equilibrium distribution can then be used to model different governing equations \cite{succi2018lattice}.  

Computationally, the classical solution procedure for Equation \ref{e:16} consists of iteratively applying two algorithmic steps, namely collision and streaming. In the \emph{collision} step, each distribution function is locally relaxed towards equilibrium,
\begin{equation}\label{e:17}
    \hat{f}_i(\mathbf{x},t)=(1-\omega)f_i(\mathbf{x},t)+ \omega f_i^{eq}(\mathbf{x},t),
\end{equation}
where $\omega=\Delta t/ \tau$. These relaxed distributions are then propagated in the direction of their corresponding velocities to the next neighboring lattice point in the \emph{streaming} step,
\begin{equation}
    f_i(\mathbf{x}+\mathbf{c}_i\Delta t, t+ \Delta t)=\hat{f}_i(\mathbf{x},t).
\end{equation}

The simplicity of these update rules, combined with the large memory footprint required for storing distribution functions, make LBM particularly appealing for quantum acceleration.
The linearity of streaming and locality of collision are especially well aligned with quantum computing paradigms and form a stark contrast to the Navier-Stokes where nonlinearity and nonlocality are directly coupled in the self advection operator. However, QCFD approaches still have to address certain challenges, such as the nonlocality of streaming and nonlinearity of collision as will be discussed shortly. Nevertheless, QLBM has emerged as one of the most actively explored frameworks in QCFD, with both theoretical developments \cite{succi2002lattice,budinski2021quantum,schalkers2024importance,steijl2022quantum} and an increasing number of tailor made software packages \cite{shinde2025utilizing,georgescu2025qlbm}.

\begin{figure}[h!]
    \centering
    \includegraphics[width=0.8\linewidth,trim=2.5cm 7.0cm 2.5cm 7cm,clip]{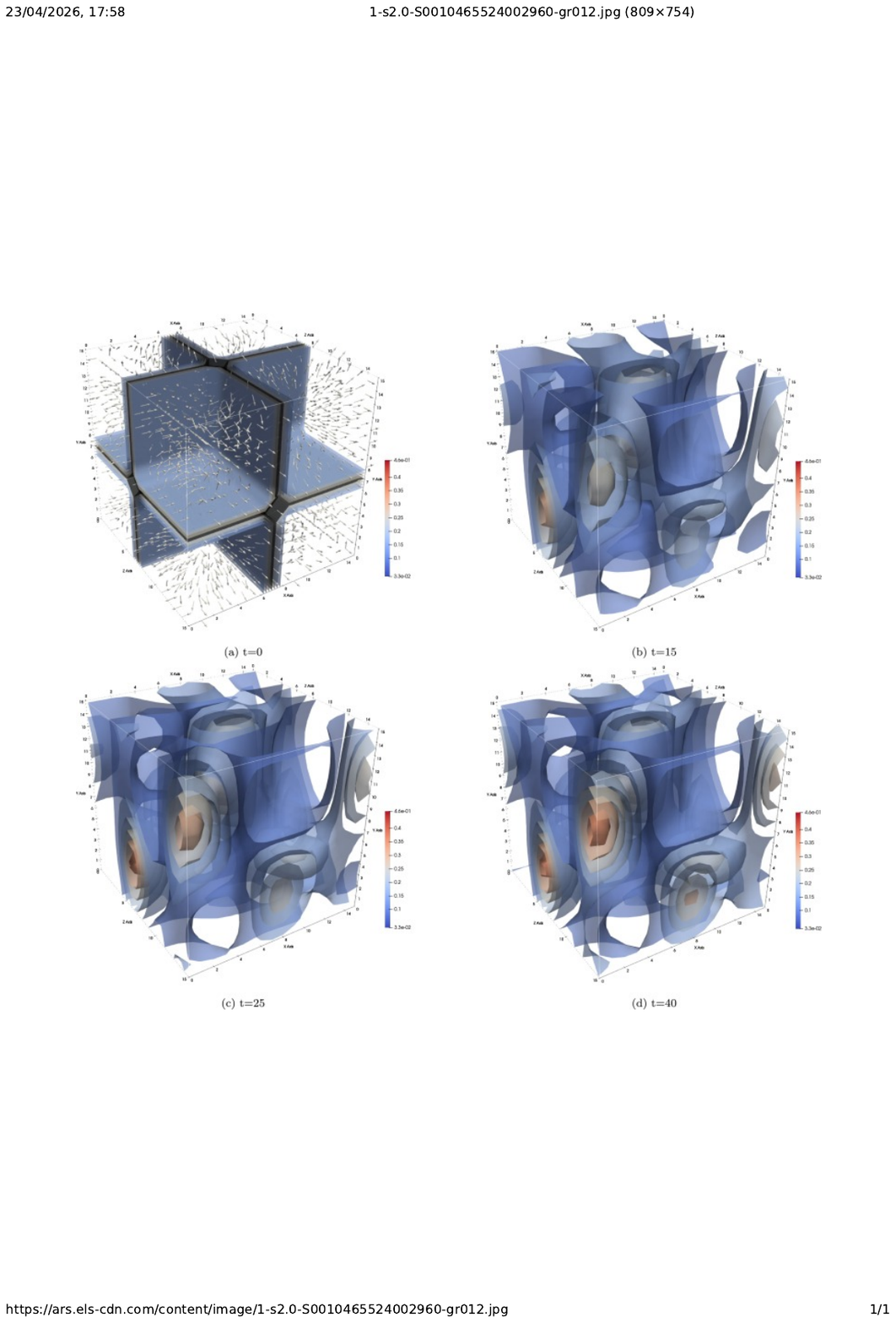}
\caption{Evolution of a Taylor--Green vortex simulated with a D3Q27 velocity set. The quantum algorithm used 17 qubits and one ancilla to represent a domain of $16^3$
 lattice cells. Only minor discrepancies were observed, most likely due to finite machine precision. Adapted from \cite{wawrzyniak2025quantum} (\href{https://creativecommons.org/licenses/by/4.0/}{CC BY}).}
    \label{fig:4}
\end{figure}

The first fully quantum LBM formulation incorporating both streaming and collision was proposed \cite{budinski2021quantum} for an advection-diffusion equation of the form
\begin{equation}
    \frac{\partial \phi }{\partial t}+\frac{\partial (u_i \phi)}{\partial x_i}=\frac{\partial}{\partial x_i}\left(D\frac{\partial \phi}{\partial x_i}\right),
\end{equation}
where $\phi$ denotes concentration, $u_i$ the advection velocity and $D$ the diffusivity. The approach was verified by simulating the movement of Gaussian hills representing transport of a conservative tracer in a channel with periodic boundary conditions, showing good agreement with classical results despite a coarse mesh. Subsequent work has built on this foundation, for instance \cite{wawrzyniak2025quantum} improved the treatment of arbitrary velocities and applied the method to higher-dimensional problems, such as the three-dimensional Taylor–Green vortex illustrated in Figure \ref{fig:4}.

A key simplification when simulating the advection-diffusion equation, relative to the full Navier-Stokes, is that the equilibrium distribution function becomes linear (albeit nonunitary)
\begin{equation}\label{e:18}
    f_i^{eq}=w_i \phi \left( 1+ \frac{c_i \cdot u_i}{c_s^2} \right),
\end{equation}
with $w_i$ velocity dependent weights and $c_s$ the speed of sound. Moreover, the macroscopic variable evaluation requires only the zero-order moment of the distribution function as 
\begin{equation}\label{e:lbm}
    \phi(\mathbf{x},t)=\sum_if_i(\mathbf{x},t).
\end{equation}

\begin{figure}[h!]
    \centering
    \includegraphics[width=0.95\linewidth,trim=1cm 3cm 5cm 6.3cm,clip]{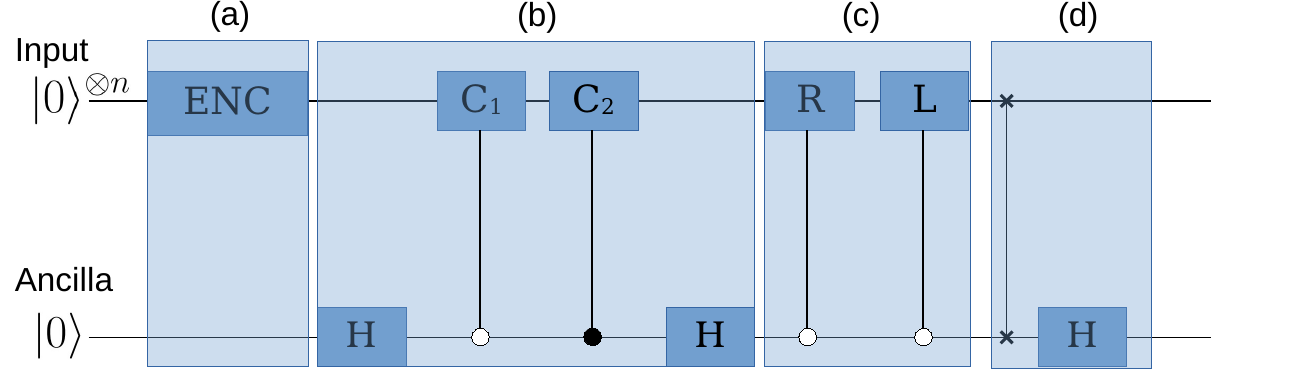}
    \caption{Quantum circuit for the D1Q2 lattice Boltzmann implementation of the one dimensional advection-diffusion equation. The circuit carries out a full time step by combining state encoding (a), the collision step via controlled operators \(C_1\) and \(C_2\) (b), propagation through the shift operators \(R\) and \(L\) (c), and final evaluation of the macroscopic variable (d).}
    \label{fig:5}
\end{figure}

The following algorithmic blocks can be readily extended to accommodate higher dimensional lattice models, but consider first the procedure for a one dimensional, two-velocity (D1Q2) model with $M$ lattice points as shown in Figure \ref{fig:5} and first described in \cite{budinski2021quantum}. Starting with flattening out the distribution function into a vector of the form  $\Phi=[\phi_{1,0} \dots \phi_{1,M-1},\phi_{2,0}\dots\phi_{2,M-1}]$, where the first index corresponds to each discrete velocity and the second refers to spatial position along the lattice.
This initial distribution function is encoded into the amplitudes of a quantum register, $R_q$, with $\log_2(2M)$ qubits and augmented by an ancilla qubit
\begin{equation}\label{e:}
    \ket{\psi_0}= \ket{0}_a\frac{1}{\left\| \phi \right\| }\sum_{j=0}^{2M-1}\phi_{i,j}\ket{j}_q,
\end{equation}
where the first qubit in $R_q$ determines the velocity direction.

In the collision step, the relaxation parameter $\omega$ in Equation \ref{e:17} is set to unity \cite{junk1999new}, corresponding to complete relaxation to equilibrium in a single time step.
This simplifies the collision operator to an update of register $R_q$ that directly enforces the equilibrium distribution of Equation \ref{e:18}.  
Formally, this update is represented by a block-diagonal matrix
\begin{equation}
   A=
\begin{bmatrix}
B_1 & 0\\
0 & B_2
\end{bmatrix},
\end{equation}
where each block $\mathbf{B}_i \in \mathbb{R}^{M\times M}$ (one per discrete velocity $c_i$) is diagonal with entries $w_i\left(1+\frac{c_i u_i}{c_s^2}\right)$. 
As this operator is not unitary it cannot be applied directly and relies on a LCU decomposition into the unitaries $C_1=A+i\sqrt{I-A^2}$ and $C_2=A-i\sqrt{I-A^2}$ such that $A=1/2(C_1+C_2)$. 
This decomposition allows enacting $A$ through the ancilla controlled sequence of operations:
\begin{equation}
    (H^\dagger \otimes I_q)(\ketbra{0}{0}_a \otimes C_1 + \ketbra{1}{1}_a \otimes C_2)(H \otimes I_q).
\end{equation}
The resultant post-collision state is
\begin{equation}
    \ket{\psi_1}=\ket{0}_a A \frac{1}{\left\| \phi \right\| }\sum_{j=0}^{2M-1}\phi_{i,j}\ket{j}_q + \ket{1}_a(i\sqrt{I-A^2})\frac{1}{\left\| \phi \right\| }\sum_{j=0}^{2M-1}\phi_{i,j}\ket{j}_q .
\end{equation}
Thus, LCU results in a probabilistic application of the desired nonunitary operator on the input state subject to successful post-measurement of the ancilla in the $\ket{0}_a$ state. Moreover, as the unitaries $C_1$ and $C_2$ are themselves diagonal with repeating entries, the physical implementation is straightforward, requiring only a single phase gate each and no off-diagonal entangling terms, which might not be true for LCU in general. 

The streaming step then propagates the relaxed distributions in the direction of their corresponding particle velocities, by shifting the amplitude coupled with basis state $\ket{i}$ to $\ket{i+1}$ ($\ket{i-1}$) for a right (left) shift. This is equivalent to a quantum random walk \cite{todorova2020quantum} conditioned on the velocity qubits in the q-register as
\begin{equation}
    \ketbra{0}{0} \otimes R_w + \ketbra{1}{1} \otimes L_w,
\end{equation}
where depending on the direction stored in the qubit the controlled operations apply a right ($R_w$) or left ($L_w$) walk implemented efficiently via a series of cascading multi-controlled NOT gates. A direct consequence is that periodic boundary conditions are enforced by default and at no additional cost, while specular-reflection boundary conditions have also been proposed \cite{todorova2020quantum} but require additional circuit overhead to keep track of solid boundaries. 
The resultant state post-streaming becomes
\begin{equation} \label{e:19}
    \ket{\psi_2}=\ket{0}_aLR\left( \frac{1}{\left\| \phi \right\| }\sum_{j=0}^{2M-1}\phi_{i,j}\ket{j}_q \right) + \ket{1}_{aq}
\end{equation}
with $\ket{1}_{aq}$ denoting the unwanted ancilla branch. 

Finally, the macroscopic variable is evaluated by summing contributions from different velocity directions at each lattice site. 
Since both distributions reside in the $\ket{0}_a$ subspace, a SWAP operation between the ancilla and velocity qubit in register $q$ first effectively reassigns one distribution to the $\ket{1}_a$ branch. 
A Hadamard gate on the ancilla then produces the sum of distributions in the $\ket{0}_a$ state and the difference in the $\ket{1}_a$ state. 
Post-selection of the $\ket{0}_a$ branch and subsequent renormalization by $2 \left\| \phi \right\|\ \sqrt{2}$ then leaves $R_q$ holding the updated concentrations ready for the next time step. 
However, note that as pointed out in \cite{xu2025improved}, successful state selection requires partial measurement of both the ancilla and additional qubits in order to filter out mixing between the $\ket{1}_{aq}$ state and $\ket{0}_a$ ancilla branch during the SWAP operation. Furthermore, the advection-diffusion equation only required evaluating the zeroth order moment, but in general macroscopic variables defined by higher order moments may also be required. For example, the first order moment (typically corresponding to fluid velocity or momentum) is given by
\begin{equation}
    u=\frac{1}{\phi}\sum_jc_jf_j,
\end{equation}
where the normalisation by the scalar field could be challenging for QLBMs \cite{wawrzyniak2025quantum} without resorting to expensive measurements. 

The main driver for a potential quantum speedup stems from the amplitude encoding of the distribution function as this allows exponential memory compression and a subsequent reduction in required qubits. However, many of the challenges faced by QLBMs also stem from this choice of encoding. 
For example, unlike in Section \ref{sec:1} where the desired information is stored in a single amplitude, or Section \ref{sec:2} that produces an entire history state in one shot, the quantum register here is iteratively evolved through time. 
Thus, if the solution at an intermediate time step is required then this is obtained via state tomography and a costly re-initialization to then continue with the simulation. Attempting this at every time step to track the temporal evolution, as is common in CFD, is clearly not viable. 
Moreover, even if the intermediate snapshots are not required, re-normalisation after ancilla post-selection requires knowledge of the state modulus which would need QST if running on actual hardware. 

Each re-initialization can be expensive, particularly as the flow field evolves and it becomes difficult to exploit any underlying structure when encoding the amplitudes. Some recent works \cite{wawrzyniak2025quantum,xu2025improved} have attempted to reduce this cost, for example by initializing the scalar variable only once instead of multiple instances across each lattice velocity direction as in Equation \ref{e:20}, which for a practical LBM simulation may contain 27 velocities (D3Q27). A reduction in circuit depth can be achieved in this way by initializing the spatial distribution of the scalar variable as 
\begin{equation}
    \ket{\Psi_0}=\frac{1}{\left\|\phi\right\|}\sum_{j=0}^{M-1}\phi_j\ket{j},
\end{equation}
before applying a Hadamard gate to the first qubit in the velocity register ($R_v$) to give
\begin{equation}
    \ket{\Psi_0}=\frac{1}{\left\|\phi\right\|\sqrt{2}}\sum_{j=0}^{M-1}\left( \ket{0}_v\phi_j\ket{j} +\ket{1}_v \phi_j\ket{j} \right),
\end{equation}
where the scalar field has ben duplicated as subspaces of the qubits in $R_v$. Similarly a Hadamard on a second qubit in $R_v$ would result in 
\begin{equation}
    \ket{\Psi_0}=\frac{1}{\left\|\phi\right\|\sqrt{2}}\sum_{j=0}^{M-1}\left( \ket{00}_v\phi_j\ket{j} +\ket{01}_v \phi_j\ket{j} + \ket{10}_v\phi_j\ket{j} +\ket{11}_v \phi_j\ket{j}  \right),
\end{equation}
where the initial distribution has been encoded four times. Notably, applying a Hadamard gate on the most significant qubit of $R_v$ duplicates the entire state, so the protocol will eventually require the use of controlled Hadamard operations to duplicate only the desired components. Consequently, different instances of the scalar field encoded in the state vector may have different normalization factors. While this must be taken into account in the collision step to make sure every field is properly rescaled, the rescaling constants can be determined \emph{a priori} for a particular circuit design and only need to be evaluated once. Moreover, rescaling of the collision matrix can be leveraged to increase the likelihood of a successful LCU post-selection by making the matrix closer to unitary \cite{wawrzyniak2025quantum}. Although uncomputation \cite{paradis2021unqomp,paetznick2014repeat} could, in principle, restore a useful pre-selection state in the case of an unsuccessful outcome, practical and general purpose recovery procedures remain difficult to implement.  

In the absence of collisions, where $\Omega_i=0$ in the discrete Boltzmann update of Equation \ref{e2222}, the dynamics reduce to pure streaming as given by
\begin{equation}
    f_i(\mathbf{x} + \mathbf{c}_i \Delta t, t + \Delta t)
    = f_i(\mathbf{x}, t).
\end{equation}
For this simplified configuration it may be possible to design the quantum algorithm with measurements delayed until the end of the time horizon \cite{todorova2020quantum} if only certain characteristics are required instead of the evolving flow field, but the collision operator is essential for wider applicability of the method. For the advection diffusion equation the only challenge associated with the operator is the nonunitarity, which can be dealt with using LCU albeit at the cost of additional runs due to its probabilistic nature. However, in more general cases (including the Navier-Stokes) the collision operator will also be nonlinear and there is no clear consensus with regards to how this should be dealt with. A common strategy is linearizing the operator using techniques such as Carlemann linearisation \cite{kowalski1991nonlinear}, which trades nonlinearity for higher dimensionality. 

The essence of Carleman linearisation is a change of variables in which a nonlinear dynamical system is mapped to a larger set of \emph{Carleman variables}, defined as monomials of the original variables. The evolution of these new variables follows directly from the original system via product calculus. This procedure yields a formally linear system of equations at the cost of an enlarged state space. Because higher-order monomials generate further monomials in a recurrent fashion, Carleman linearisation produces an infinite-dimensional linear system, which must be truncated (or closed) to remain computationally tractable even if such truncation inevitably introduces approximation error.
Carleman linearisation for the LBM in a classical context \cite{itani2022analysis} is known to face two significant challenges, firstly a combinatorial increase in the number of variables leads to a blow up of degrees of freedom. Secondly, the approach leads to a coupling between the linearized collision term and the the streaming step, essentially exchanging local nonlinearity for nonlocal linearity and turning a nonlinear one-body problem into a linear k-body problem, where k is the truncation order. Consequently, streaming is also represented by an infinite dimensional system which if truncated results in the step no longer being exact.   

Interestingly, the loss of exact streaming, a key advantage of LBM, can be avoided when applying the technique in a quantum paradigm and the large Hilbert space can be used to embed the Carleman variables as eigenvalues of operators compactly. This was demonstrated in a recent work \cite{itani2024quantum} which derived unitary collision and streaming operators using Carlemann linearisation to truncate encoding in the bosonic Fock space. 
However, their collision operator complexity scales as a power of the number of timesteps due to the linear embedding induced nonlocality, making it uncertain if a quantum advantage could be retained. 
Furthermore, the traditional unitary streaming conventionally applied to a binary encoding of position \cite{todorova2020quantum,budinski2021quantum} is incompatible with their linear embedding of the collision operator without mixing states in time. 
While the authors in \cite{itani2024quantum} do suggest alternative streaming operators, albeit at the cost of increased complexity, the idea that certain encodings might preclude the potential for unitary streaming \emph{and} collisions was also explored in \cite{schalkers2024importance} and highlights the the subtle trade-offs that occur between the advantages and disadvantages of a particular encoding strategy.

A potential approach \cite{sanavio2024lattice} is to relax the requirement of completely unitary evolution, fully encode the Carleman variables into an amplitude basis, coupled with unitary streaming and a nonunitary collision operator applied probabilistically. 
Yet, the issue of nonlocality cannot be avoided and as the collision operator is no longer local it cannot be written in the standard sparse, block diagonal form that is amenable for efficient gate implementation \cite{jordan2009efficient}. 
As such the required number of two-qubit gates even for a second order Carleman truncation scales as $\mathcal{O}(N^4Q_v^4)$ for the number of lattice points ($N$) and discrete velocity directions ($Q_v$) which is well beyond the scope of any current quantum hardware by several orders of magnitude. 
Thus, despite the promising convergence properties of Carlemann linearisation applied to LBM \cite{li2025potential,itani2022analysis,sanavio2024three} in terms of truncation order, the approach quickly runs into an unfeasible circuit depth barrier even for a small number of grid points. 
Although the nonlocal terms can be minimised by iterating the quantum algorithm only a single time step for each run, thereby fixing the circuit depth (still in the order of tens of thousands) independently of the number of lattice points \cite{sanavio2024lattice}, this merely shifts the burden to the repeated initialization and readout steps, which are significant encoding challenges in their own right.

Even though substantial effort has been devoted to incorporating nonlinear collision operators within amplitude encoded frameworks, it is clear that there are still many important challenges to overcome. However, an alternative approach \cite{steijl2022quantum,steijl2022quantumm,moawad2022investigating} is to employ a basis encoding, in which fluid variables are represented using a reduced-precision floating-point format tailored for quantum circuits. 
This representation uses a small number of mantissa qubits together with an asymmetric exponent bias to compensate for the limited number of available qubits, motivated by the fact that QLBM variables are typically normalized by the lattice speed of sound and remain close to zero.
Within this framework, modular quantum arithmetic circuits for addition, multiplication, and squaring can be composed to directly evaluate the nonlinear collision operator, although due care must be given to include correct rounding functionality \cite{steijl2022quantum}, even at the cost of significantly increased circuit complexity. 
Although this enables arithmetic operations that are difficult in amplitude encoding, the potential source of quantum advantage over classical methods is not immediately as clear.
One suggestion \cite{steijl2022quantumm} has been to apply basis-encoded evaluation of the collision operator only locally within the algorithm, while retaining amplitude encoding for the remaining steps. 
This hybrid approach, however, would necessitate repeated conversions between encodings (see Section \ref{s:0}), making it uncertain whether the associated overhead would permit any meaningful speedup in practice.

Moreover, the implications of choice of encoding extend to include the streaming step as well. While the streaming increment/decrement in physical space has been commonly employed in QLBMs successfully, it is also possible \cite{schalkers2024efficient} to use a modified Quantum Draper Adder \cite{draper2000addition} for carrying out the addition in Fourier space. 
This leads to a more efficient circuit implementation with a significant reduction in the required number of CNOTS.
Similarly, encoding the velocity directions using a unary encoding is less efficient in the number of qubits due to only linear scaling of valid configurations, but results in streaming operators that are always controlled on just a single direction qubit. 
This results \cite{tiwari2025algorithmic} in a significant reduction in the number of CNOT gates, suggesting that the increase in qubit count might be a good investment in return for more efficient circuits and highlighting subtle tradeoffs that should be considered when deciding on an encoding protocol.

Additionally, incorporating meaningful boundary conditions requires efficient strategies for encoding solid objects within the QLBM framework.  
Specular reflection boundaries, for instance, rely on identifying particles that have streamed into the interior of a solid and redirecting them back into the fluid domain before the next streaming step.
This is not trivial in a quantum setting as directly adjusting the position of particles based on their current position constitutes operations on qubits controlled by their own states, which would be nonunitary. 
For simple geometries it may suffice to introduce some ancilla boundary condition qubits to keep track of which states represent a wall collision and subsequently reflecting the normal velocity component, but this can run into problems particularly around corners. A recent work \cite{schalkers2024efficient} introduced a novel object encoding protocol using a quantum comparator, resilient to edge cases such as corner boundaries and was coupled with a modified velocity mapping so that the velocity could be reflected with a single qubit flip. Moreover, there is evidently a complex interplay between encoding multiple quantities, such as velocities and boundaries beyond simply periodic conditions, that is commonly overlooked.  

In summary, QLBMs demonstrate that encoding is a critical design choice with inherent trade-offs. No single protocol is optimal across all regimes, while some prioritize quantum parallelism and circuit economy, others emphasize numerical stability, error resilience, or versatility across models and boundary conditions. Crucially, researchers should distinguish challenges intrinsic to the governing equations/discretization from those introduced by the encoding itself.
In addition to designing isolated, individual quantum primitives, a more holistic perspective is needed where encoding choices should be reconsidered in the context of the complete algorithmic pipeline, recognizing that different components may impose distinct requirements on the encoding strategy.

A detailed discussion of the inherent limitations of LBM discretisation lies beyond the scope of this work, but it is useful to highlight how such issues may shape future QLBM developments. Prior studies \cite{chopard2009lattice} have shown that the LBM approximation commonly employed and shown here does not exactly recover the advection-diffusion equation in the macroscopic limit, but instead introduces a spurious error term. One classical remedy is to add a simple body-force correction at the LBM level, which cancels this error in the continuum description. However, how such a forcing scheme could be encoded consistently within the QLBM remains unclear. 
In addition, it is known that refining the spatial grid while keeping the time discretisation fixed can degrade accuracy rather than improve it. Addressing this requires the ability to handle variable relaxation times, which therefore represents an important open challenge for future QLBM research. \cite{wawrzyniak2025quantum}. 
\section{Summary And Outlook}\label{A:4}
Quantum computing has the potential to accelerate CFD, but practical impact depends heavily on choices made at the level of encoding data. Beyond the familiar challenges of loading classical data into quantum states and extracting observables, encoding decisions can shape how boundary conditions are imposed, how nonlinear terms are represented and how coupled equations interact. These considerations can be subtle, especially for classical domain specialists who are new to the field, so our goal is to make the trade-offs explicit, helping practitioners select encodings that align with their particular requirements.

We surveyed several widely used quantum approaches and examined, in detail, the nuances of their encoding choices in terms of advantages, challenges and potential opportunities. Section \ref{sec:1} addresses the measurement bottleneck in amplitude encoding and outlines strategies for extracting observables efficiently. Section \ref{sec:2} describes how nonlinearity can be simulated via interacting copies in amplitude encoded formulations. Section \ref{sec:3} turns to quantum annealing and basis encodings, showing that forgoing the exponential compression of amplitude encoding need not be detrimental and alternative encodings can still deliver practical advantages. Finally, Section~\ref{sec:4} illustrates how encoding decisions interact with multiple aspects of the algorithm such streaming, collisions, boundary conditions and readout, making it essential to disentangle limitations inherent to the physical problem from those introduced by the encoding selected at the outset.

The methods surveyed here are not exhaustive as QCFD is a broad and rapidly evolving field. Variational approaches, for example, recast differential equations as minimization problems in which a parametrized quantum circuit evaluates a loss function while a classical optimizer updates the parameters (see \cite{tennie_2024} for applications to nonlinear PDEs and their hardware implications). One line of work \cite{lubasch2020variational} employs tensor networks over multiple amplitude-encoded copies with an ancilla-mediated, nonlocal loss measurement offering exponential memory compression but lacking convergence guarantees. An alternative \cite{kyriienko2021solving} encodes data via single qubit rotations and fits solutions with a finite Chebyshev basis but analytic bounds on optimization cost and convergence remain open. Quantum reservoir computing \cite{pfeffer2022hybrid} replaces the classical reservoir with a high-dimensional quantum state, enabling time-series prediction for low-dimensional nonlinear systems but incurring heavy sampling costs.

We also do not cover quantum lattice gas automata, historically associated with Type-II quantum architectures (i.e. a large network of small quantum computers interconnected via classical channels) \cite{yepez2001quantum} but recently revisited with new formulations \cite{kocherla2024fully,zamora2025efficient}, or generalized (inverse) Madelung transforms that recast fluid dynamics into modified Schr\"odinger type equations. While a quantum-fluid correspondence has long been noted, the Madelung fluid differs from Navier-Stokes in dissipation, non-gradient flows, and the role of the quantum potential. However recent work \cite{salasnich2024quantum,meng2023quantum} proposes potential solutions for each of these challenges. Across all these avenues, encoding choices introduce specific caveats that must be stated explicitly when assessing practicality.

\begin{figure*}[h!]
\centering
\includegraphics[width=0.95\textwidth,trim=4cm 0cm 3cm 0cm,clip]{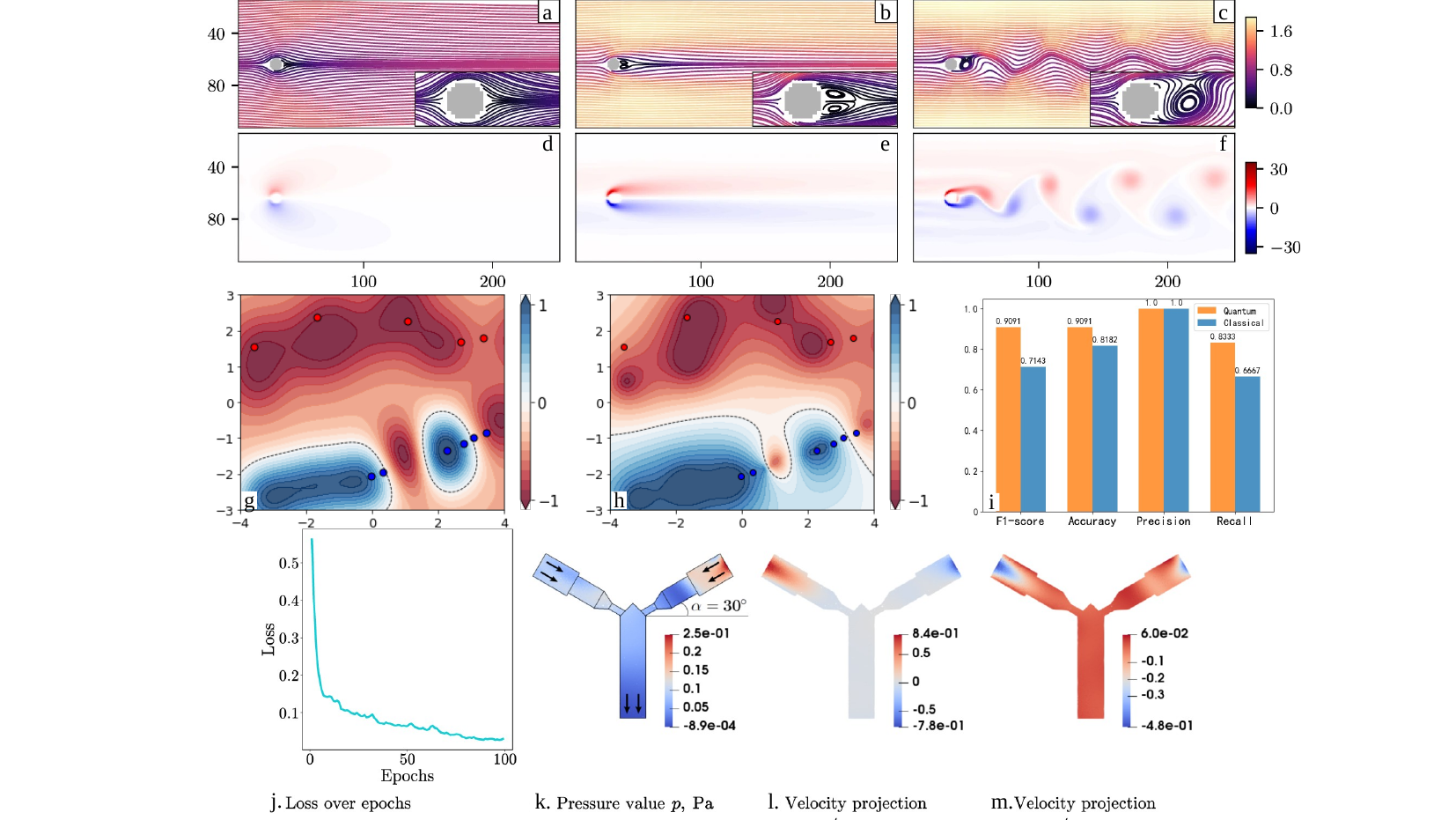}
\caption{Illustrative examples of application-level results in quantum and quantum-inspired fluid modelling. Rows 1 and 2 show a quantum-inspired matrix-product-state formulation applied to flow around a cylinder at increasing Reynolds number from left to right, with velocity streamlines in the top row and vorticity in the bottom row. As the Reynolds number increases, the onset of the classical K\'arm\'an vortex street is observed in the wake of the cylinder. Adapted from \cite{peddinti2024quantum}. Row 3 shows a quantum support vector machine for flow classification around an airfoil, including the decision boundaries for the classical (g) and quantum (h) cases, together with a performance comparison (i), where the quantum model outperforms the classical counterpart across several metrics. Adapted from \cite{yuan2023quantum}. Row 4 shows a hybrid physics-informed neural network applied to flow in three-dimensional Y-mixers where the observed asymmetry between the left and right branches was suggested to arise from the data encoding strategy. Adapted from \cite{sedykh2024hybrid}. All figures licensed under \href{https://creativecommons.org/licenses/by/4.0/}{CC BY}.}
\label{fig:6}
\end{figure*}

At the same time, it is important to emphasize that the field is not defined solely by its open challenges. A growing body of work has already reported encouraging results across a range of quantum and quantum-inspired algorithms applied to physically meaningful problems. Figure~\ref{fig:6} highlights a small selection of these advances, including a matrix-product-state formulation with logarithmic complexity in mesh size that can also accommodate complex boundaries \cite{peddinti2024quantum}, a quantum support vector machine for flow classification around objects such as airfoils that outperforms its classical counterpart \cite{yuan2023quantum}, and a hybrid quantum physics-informed neural network for three-dimensional Y-shaped mixers that achieves substantially greater accuracy than a purely classical implementation \cite{sedykh2024hybrid}. This selection is by no means comprehensive, but instead serves to illustrate the breadth of applications in which quantum computing may prove useful.

Beyond the current theoretical developments being put forwards, algorithms must ultimately be realized on actual quantum hardware. Because architectures differ in qubit connectivity, native gates, coherence, precision, and readout, no single platform will suit every QCFD task. This makes encoding strategy a hardware aware design choice, requiring explicit co-design between the problem formulation and the target device in addition to purely algorithmic considerations.
We therefore advocate treating encoding as an explicit design variable that is revisited early and iteratively, rather than a fixed assumption. Continued progress will likely depend on systematic comparisons of alternative encodings, the development of hybrid or adaptive schemes, and ultimately the formulation of new encoding paradigms tailored to the specific demands of the problem at hand.

\newpage
\section*{Acknowledgments}
The authors acknowledge funding through UKRI EPSRC projects (EP/W00772X/2 and EP/Y004515/1).
\bibliography{references}

\end{document}